\documentclass[aps,twocolumn,notitlepage,showpacs]{revtex4-1}

\usepackage{epsfig}

\usepackage{amsmath}
\usepackage{braket}
\usepackage{graphicx}
\usepackage{wrapfig}
\usepackage{verbatim}
\usepackage{setspace}
\usepackage{indentfirst}
\usepackage{feynmf}
\usepackage{color}
\usepackage{booktabs}
\usepackage{multirow}
\usepackage{subfigure}
\usepackage{fancybox}
\usepackage{comment}

\def\be{\begin{eqnarray}}\def\ba{\begin{eqnarray}}
\def\ee{\end{eqnarray}}\def\ea{\end{eqnarray}}
\def\ben{\begin{enumerate}}\def\bitem{\begin{itemize}}
\def\een{\end{enumerate}}\def\eitem{\end{itemize}}
\def\no{\nonumber\\}
\def\la{\langle}\def\ra{\rangle}

\begin{document}

\author{Y. Kim}
\affiliation{Rare Isotope Science Project, Institute for Basic Science, Daejeon 305-811, Republic of Korea}

\author{P. Papakonstantinou}
\affiliation{Rare Isotope Science Project, Institute for Basic Science, Daejeon 305-811, Republic of Korea}

\title{Proton pygmy resonances: predictions for $N=20$ isotones}

\begin{abstract}
We study theoretically the low-energy electric-dipole response of $N=20$ isotones.   
We present results from a quasiparticle random-phase approximation (QRPA) and a continuum random-phase approximation (CRPA), 
and we compare them with results for the mirror $Z=20$ nuclei. 
According to our analysis, enhanced $E1$ strength is expected energetically well below the giant dipole resonance in the proton-rich isotones. 
Large amounts of $E1$ strength in the asymmetric $N=20$ isotones are predicted, unlike their equally asymmetric $Z=20$ mirror nuclei, pointing unambiguously to the role of structural effects such as loose binding. 
A proton-skin oscillation could develop especially in $^{46}$Fe. 
The proper description of non localized threshold transitions and the nucleon effective mass in mean-field treatments 
may affect theoretical predictions. 
We call for systematic theoretical investigations to quantify the role bulk-matter properties, in anticipation of  measurements of $E1$ transitions in proton-rich nuclei.

\end{abstract}

\maketitle

\section{Introduction\label{Sec:intro}}

The low-energy electric-dipole transitions of stable and unstable nuclei have been under investigation for more than two decades.
The effort is posed to continue, as new experiments on exotic nuclei are being analyzed and new facilities with improved capabilities are being developed.
In the future the theoretical feedback may include results from methods drawing on {\em ab initio} nuclear structure, namely the use of microscopic nuclear interactions.
The concerted effort promises not only valuable empirical input into nucleosynthesis studies, but also crucial discoveries regarding nuclear structure far from the stability valley, and the nuclear equation of state.

The origin of low-energy $E1$ strength remains under debate, even in the case of stable nuclei.
A particularly vexing question is whether or not highly asymmetric nuclei, with an excess of neutrons (or protons), can develop a collective dipole vibration, whereby
the neutron (or proton) skin oscillates against an (approximately) iso-symmetric core.
Such a collective resonance is usually called a pygmy resonance. 
Alternatively, the question formulated in configuration language is whether a collective resonance develops, through coherent excitations of many valence neutrons (or protons). 
The answer holds particular relevance for validating the predictive power and limitations of nuclear structure models in unexplored regions of the nuclear chart.
It is related to predictions about shell structure, the softness of the symmetry energy, phonon coupling, and broad resonances in the continuum.
The reader may consult Refs.~\cite{PVK2007,KrS2009,SAZ2013} for reviews on the status of the above issues in the recent past.

The majority of studies devoted to the above issues, both theoretical and experimental, have focused on neutron-rich nuclei.
One reason is that the highly accessible heavy stable nuclei are, in an obvious sense, neutron-rich.
The other side of the stability valley, comprising proton-rich nuclei, is no less interesting and challenging.
Because the $N=Z$ line reaches the proton drip line by $Z\approx 50$, nuclei with a few excess protons ($T_z<-3$) reach up to medium mass only.
Their structure is determined to a large extent by the Coulomb field, besides the symmetry energy, and of course by shell effects, pairing, and their proximity to the drip line.

Early theoretical work, within the relativistic self-consistent quasi-particle random-phase approximation (RQRPA), predicted the development of a proton-skin oscillation in proton-rich $N=20$ isotones and very neutron-deficient Ar isotopes~\cite{PhysRevLett.94.182501}.
The fragmentation of such a resonance was further studied in Ref.~\cite{PhysRevC.77.024304} using a unitarily transformed Argonne V18 potential.
Within the relativistic model, neutron-deficient Ar isotopes were predicted to develop a halo-like structure~\cite{PVK2007}, characterized by a diffuse proton distribution, favoring soft dipole excitations.
The somewhat heavier $N=20$ isotones, on the other hand, were predicted to develop less diffuse proton skins.
Results on $^{48}$Ni within the Skyrme-continuum random phase approximation predicted a similar amount of strength as RQRPA~\cite{PPP2005}, but the origin could be single-particle excitations into the continuum.
Studies of light Ne and F isotopes, within the RQRPA~\cite{PhysRevC.85.044307} and the shell model~\cite{MaT2011}, predict an exotic proton-dominated transition in $^{17}$Ne. 
A Gogny-QRPA model preticts a collective low-lying proton excitation in $^{18}$Ne~\cite{MPD2011}. 
Large-scale continuum quasiparticle random-phase approximation (CQRPA) calculations predict up to five percentage points of the $E1$ energy-weighted sum to be exhausted by transitions below 10~MeV in various proton-rich nuclei~\cite{DaG2012}. 
Recent measurements in $^{32,34}$Ar isotopes~\cite{Lep2013} will help assess some of the above predictions. 

In the present theoretical study we focus on $N=20$ isotones.
These nuclei are heavy enough to present proton-skin candidates (as opposed to proton-halo nuclei).
The $N=20$ isotones include the stable and symmetric $^{40}$Ca, and the magic drip-line nucleus $^{48}$Ni.
The mirror nucleus of the latter, namely $^{48}$Ca, is stable and thoroughly studied and can be used to pinpoint the role of the nucleon separation energy.

One of our main motivations is that proton-rich nuclei in this mass region could be accessible in future radioactive-ion beam facilities. 
Furthermore, as a testing ground, they can help us address a number of questions which have risen in past investigations.
The questions include, for example, 
the degree to which covariant and non-relativistic models 
agree in the case of proton-rich nuclei~\cite{PPP2005},
and
the role of shell effects and the continuum in determining the strength distribution (cf. a related study of Ni isotopes~\cite{PHR2015} and the thorough investigation of Ref.~\cite{CDA2013}).
In the present work we use two non-relativistic models with complementary capabilities: the self-consistent quasiparticle random-phase approximation (QRPA) with the Gogny D1S interaction and the continuum random phase approximation (CRPA) with Skyrme interactions.
The former includes pairing effects, while the latter offers a proper treatment of the continuum.

Our article is organized as follows. 
In Sec.~\ref{Sec:Theory} we introduce the theoretical models and the quantities of interest.
In Sec.~\ref{Sec:Results} we present our results for the $N=20$ isotones.
We summarize our results and conclusions in Sec.~\ref{Sec:Concl}.

\section{Theoretical framework}\label{Sec:Theory}
Here, we briefly discuss the models  used in the present study and essential physical quantities for our analysis.

For a study of weakly bound nuclei far from the valley of stability, it is indispensable to consider pairing correlations.
A self-consistent approach of mean field and pairing correlations is the Hartree-Fock-Bogoliubov (HFB) theory which has 
 been successfully applied  not only in stable nuclei, but also in exotic nuclei with a large proton or neutron excess.
In the QRPA  the input ground states of (open-shell) nuclei are from the HFB theory in the canonical basis,
in which HFB ground states can be expressed in terms of the BCS variational parameters and in which the single particle density matrix is diagonal.
In this study we work with the self-consistent quasiparticle random-phase approximation (QRPA) with the Gogny D1S interaction~\cite{BGG1991} (QRPA+D1S). 
The latter has provided interesting results for the electric-dipole response in stable and neutron-rich nuclei~\cite{PPR2011,PHP2012,PHP2014,Der2014,PHR2015} 
which compare well with existing data.  

Despite its great success in describing collective excitations of stable and unstable nuclei, the QRPA fails to treat highly excited states with
continuous spectra  properly since typical QRPA calculations are based on discrete spectra. 
This issue is especially relevant for very weakly bound systems such as the proton-rich ones being studied here. 
To deal with the coupling between bound states and (quasi)particles in the continuum such as a two-particle configuration with one nucleon in the discrete bound level and the other in the continuum, one needs to construct the (Q)RPA with the appropriate boundary conditions and, ideally, starting from the ground state of
a nucleus calculated using couplings to the continuum. 
A continuum RPA (CRPA) without pairing correlations, based on the Skyrme-Hartree-Fock calculations of the nuclear ground state~\cite{ReXX}
is employed in \cite{PPP2005}. The same CRPA with the SLy4 Skyrme interaction will be employed in this work (CRPA+SLy4). 
The SLy4 functional is chosen because it predicts similar nuclear-matter properties of possible relevance as the D1S functional, namely   
symmetry energy ($J=32,31$~MeV for SLy4, D1S, respectively) and nucleon effective mass ($m^{\ast}=0.7m$) and they are both asy-soft~\cite{Che2012}. 
The difference in the Thomas-Reiche-Kuhn (TRK) enhancement factor (0.25 in SLy4 but 0.6 in D1S) is expected to 
influence the peak energy of the giant dipole resonance (GDR)~\cite{PBB2005}, but not the properties of pygmy dipole strength~\cite{Rei1999,ReN2010,ReN2013}. 

We will also discuss results with other Skyrme parameterisations. 

The single-particle basis in which the HFB equations are solved consists of 15 harmonic-oscillator shells in this work. 
The length parameter $b_{\mathrm{HO}}$ is such that the ground-state energy is minimized.  
The QRPA equations are formulated in the HFB canonical basis. 
By solving the QRPA equations, we can obtain the excitation energy $E_\nu$ and the two-quasiparticle amplitudes, 
$X^\nu$ and $Y^\nu$, of the $\nu$-th QRPA state. 
From these quantities we can calculate the transition probabilities and all other quantities of interest. 
For details on the QRPA implementation and its performance the reader may consult Ref.~\cite{HPR2011}. 

The CRPA method is formulated in coordinate space using Green's functions. 
First the Skyrme-Hartree-Fock (SHF) equations are solved to obtain the single-particle potentials and densities. 
The unperturbed $ph$ propagator $G^0$ is then evaluated, as well as the residual $ph$ interaction $V_{\mathrm{res}}$, in matrix form, 
where the rows and columns correspond to radial points on a mesh, $(r,r')$, and different blocks correspond to different operators in the residual interaction 
and the two possible values of isospin $t$. 
The RPA propagator $G^{\mathrm{RPA}}$ is then calculated from the equation 
\begin{equation} 
G^{\mathrm{RPA}} = (1 + G^0 V_{\mathrm{res}})^{-1}G^0 
\end{equation} 
by matrix inversion. 
We use a radial mesh of 16~fm for $^{40,48}$Ca and 18~fm for $^{48}$Ni and a step size of 0.08~fm. 
 
In this work only spin-independent terms of the residual interaction is considered in CRPA. 
Therefore, our CRPA implementation is not fully self-consistent. 
To correct for the missing effects we scale the residual interaction by a factor, such that the spurious state is obtained at zero energy. 
The required factors for $^{40}$Ca, $^{48}$Ni, $^{48}$Ca equal 1.03, 1.11, 1.14, respectively, when the SLy4 functional is used. 

To investigate the properties of $N=20$ isotones (and $^{48}$Ca) we evaluate several quantities of interest using the two models.
A quantity of primary interest is the transition probability from the (QRPA/CRPA) excited state to the ground state,
a measure for the dynamical response of a nucleus to an external operator $Q$. 
In our case this will be an isoscalar (IS) or isovector (IV) dipole operator.
The IS and IV electric dipole operators are given by 
\ba
&&Q_{\rm IS}=e\sum\limits_{i=1}^{A}\left( r_i^3 -\frac{5}{3} \la r^2\ra r_i  \right) \sqrt{3} Y_{1M}({\hat{\bf r}_i})\, , \no
&&Q_{E1}=e\frac{N}{A}  \sum\limits_{p=1}^{Z} r_p Y_{1M}(\hat{\bf r}_p)-e\frac{Z}{A}\sum\limits_{n=1}^{N}r_n Y_{1M}({\hat{\bf r}_n}) 
\label{Eq:E1op} 
\ea
in standard notation (see, e.g., Ref.~\cite{HPR2011}). 
In the framework of our self-consistent QRPA method the above $E1$ operator is equivalent to its uncorrected form,  $Q_{E1}'=e\sum_{p=1}^{Z} r_p Y_{1M}(\hat{\bf r}_p)$, in which only protons contribute. 

The transition matrix element of the electric dipole operator with spherical symmetry in the QRPA reads
\ba
b_{qq^\prime}^\nu=(X_{qq^\prime}^\nu -Y_{qq^\prime}^\nu) (v_q u_{q^\prime} -v_{q^\prime} u_q)\la q| Q|q^\prime\ra\, ,
\label{Eq:bqqn} 
\ea
where $u$ and $v$ are the occupation probability factors of  single-nucleon states in standard notation.
The transition strength $B(E_\nu)$ is then given by
\ba
B(E_\nu)=\left| \sum\limits_{q<q^\prime} b_{qq^\prime}^\nu \right|^2
. 
\ea
A smooth transition-strength distribution can be obtained by folding with a Lorentzian, 
\ba
R(E)=\sum\limits_{i} \frac{B(E_i)}{\pi}\frac{\Gamma}{(E-E_i)^2+\Gamma^2}
. 
\label{Eq:Lor}  
\ea
Here, $2\Gamma$ is the desired full width at half maximum (FWHM) for each isolated transition. 

We note that, in existing studies using the  QRPA+D1S model~\cite{PHP2012,Der2014,PHP2014}, it was found possible to reproduce the measured $E1$ strength in various nuclei by considering a 3~MeV downward shift, i.e., by accepting a global discrepancy of 3~MeV in the low-energy region between the model and the data. 
The origin of this systematic discrepancy is not clear, but it may lie in convergence issues or the need to include effects beyond RPA (cf. Refs.~\cite{GoK2002,CDA2013}). 
In the discussion of our results it is made explicit when we take into account this empirical shift.

In the CRPA the strength distribution is determined by the polarization diagram via the Green's function 
\begin{equation} 
R(E) = \frac{dB(E)}{dE} = -\frac{\Im}{\pi} 
\sum_{t,t'} 
\int d^3 r d^3 r' 
f_t(\vec{r}) 
G_{tt'}(\vec{r},\vec{r'};E) 
f_{t'}(\vec{r}') 
,
\end{equation}  
where $f_t(r)$ 
denotes the spatial and isospin dependence of the external field.  
The obtained $R(E)$ automatically includes the escape width. 
The distribution can be smoothed further for presentation purposes by adding an imaginary part to the $ph$ energy in $G^0$, 
analogous to the parameter $\Gamma$ in the Lorentzian of Eq.~(\ref{Eq:Lor}). 

The transition density is also an essential quantity to study the dynamics of nuclear collective motion, and it is defined as the
transition matrix element of the density operator
\begin{equation}
\delta\rho^\nu (\bf r)=\la\nu| \sum\limits_{i} \delta(\bf r - \bf r_i)   |0\ra\, .
\end{equation}
In the QRPA, the radial part of the transition density in the $1^-$ channel can be expressed as
\ba
\delta\rho_t^\nu = \!\!\sum\limits_{q<q^\prime, t_q=t_q^\prime} \!\! (X_{qq^\prime}^\nu-Y_{qq^\prime}^\nu)
(v_qu_q^\prime -v_q^\prime u_q)R_q(r)R_{q^\prime}(r) \, ,
\ea
where $t$ is for the isospin (neutron or proton) and $R(r)$ is the radial wave function of the QRPA state.
In the CRPA model, transition densities can be evaluated as follows. 
Near a resonance of interest, at energy $E$, the radial Green's function is proportional to a product of radial transition densities~\cite{BeT1974}, 
\begin{equation} 
\Im G_{t,t'}(r,r';E) \propto \delta \rho_t (r;E) \delta \rho_{t'}(r';E)  
. 
\label{Eq:factorize} 
\end{equation} 
Folding the above quantity with the radial part of the external field, $f_t(r)$, we get a function  
\begin{equation} 
\delta \tilde\rho^{(f)}_t(r;E) = \sum_{t'}\int \Im G_{tt'}(r,r';E) f_{t'}(r') dr' 
, 
\end{equation} 
which is proportional to $\delta \rho_t (r;E)$, if Eq.~(\ref{Eq:factorize}) holds. 
If we now fold this with $f_t(r)$, we obtain $\pi$ times the transition strength per energy unit. 
Although the above factorization is not always perfect, the procedure helps us to obtain a quantity, 
loosely called a transition density, which determines how the nuclear density couples to the external field at a given energy. 

The last quantity that we will examine in this work is the energy weighted-sum (EWS) of $E1$ strength. 
In QRPA we may write it as 
\ba
m_1(E_j)=\sum\limits_{i:E_i<E_j} E_i\cdot B(E_i)  \, , 
\ea
By $m_1^{\rm tot}$ we will denote the sum over all the available QRPA states.
In CRPA the equivalent quantities are obtained by an energy-weighted integration of the transition strength probability. 

\section{Results}\label{Sec:Results}


From the shell-structure point of view, the $N=20$ isotones differ only in the occupation number of the proton $0f_{7/2}$ orbital, 
which is zero for $^{40}$Ca and 8 for $^{48}$Ni. 
The orbital can be spatially extended if its binding energy is small. 
According to the HFB model with the D1S functional, that energy is lower than 1~MeV in $^{46}$Fe and $^{48}$Ni. 
According to the SHF model with the SLy4 functional, the orbital is somewhat unbound in $^{48}$Ni, but localized via the Coulomb barrier. 
This is in contrast to the well-bound neutron $0f_{7/2}$ orbital in all mirror nuclei. 
Accordingly, the proton separation energies in the $N=20$ isotones are quite different from the neutron separation energies in the mirror Ca isotopes. 
From the binding energies tabulated in the 2012 evaluation~\cite{AME2012} we find that the neutron separation energies in the Ca isotopes well exceed 9MeV. 
For the $Z=20,22,\ldots,28$ isotones, on the other hand, the proton separation energies take the values $8.32,~3.75,~3.03,~1.54,~0.692$MeV, respectively. 

\begin{figure}   
\includegraphics[width=7.5cm]{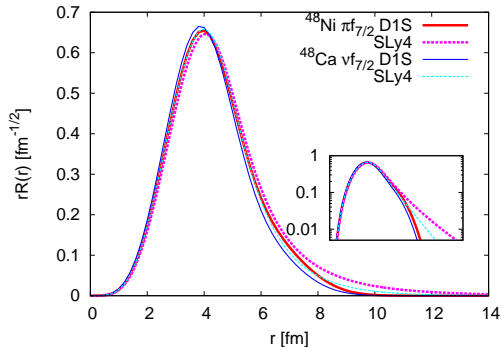}
\vspace*{-1.0cm}
\caption{(Color online) Radial wavefunction of the proton $0f_{7/2}$ orbital in $^{48}$Ni 
and of the neutron $0f_{7/2}$ orbital in $^{48}$Ca according to the Hartree-Fock model with the SLy4 functional (solved in coordinate space) and the D1S functional (in a harmonic-oscillator basis). 
The inset shows the same in semi-logarithmic scale.  
\label{fig0}}
\end{figure}

In Fig.~\ref{fig0} we show the corresponding radial wavefunctions in $^{48}$Ni as well as the neutron $0f_{7/2}$ wavefunction of the mirror nucleus $^{48}$Ca according to both models. 
%
%
Both models produce a more-extended proton wavefunction in $^{48}$Ni than the (much more bound) mirror neutron wavefunction. 
This result is dictated by basic quantum mechanics as it originates in the much weaker binding of the orbital in the proton-rich counterpart than the neutron-rich one.
Also visible is the different asymptotic behavior predicted by the two models. 
The extended tail of the wavefunctions produced by the Hartree-Fock model solved in coordinate space is absent form the results of the model 
solved in the harmonic-oscillator basis. 

The difference between the proton and neutron root mean square radius, $R_p-R_n$, defines the thickness of the proton skin. The absolute values obtained with the HFB+D1S model are listed in Table~\ref{table1}. 
Also listed are the absolute values for $^{46,48}$Ca, namely the mirror nuclei of $^{46}$Fe and $^{48}$Ni, corresponding to the neutron-skin thickness. 
The values $|R_p-R_n|$ for $^{40}$Ca, $^{48}$Ca and $^{48}$Ni obtained with the SHF+SLy4 model are, respectively, 
0.076, 0.111 and 0.281~fm. 
We observe that the proton skins of the proton-rich nuclei are much thicker than the neutron skins of the mirror nuclei. Therefore, even though no skin-oscillation has been observed in $^{48}$Ca at low energies~\cite{Der2014}, it remains possible that a proton-pygmy mode develops in $^{48}$Ni or $^{46}$Fe -- a possibility which we aim to assess in this work. 

\begin{figure}   
\hspace*{-0.7cm}
\includegraphics[width=9.3cm,height=9cm]{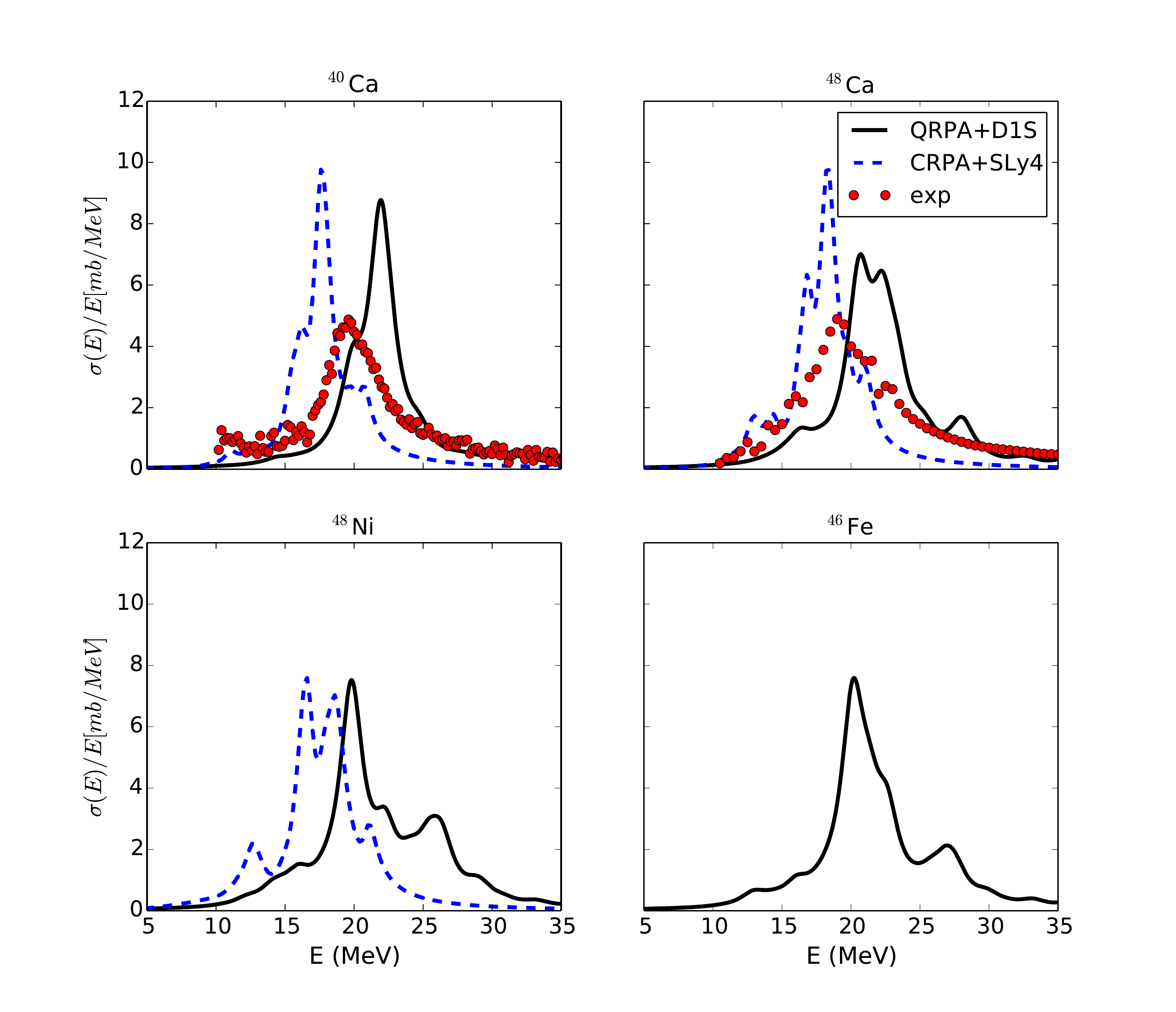}
\caption{(Color online) Isovector dipole strength distributions smoothed with a Lorentzian of $\Gamma=1$ and experimental data where available. }
\label{fig1}
\end{figure}

In Fig. \ref{fig1} we compare the GDRs obtained from the two models and from experiments~\cite{CDFE}, if available. 
The CRPA model can be applied for closed-shell configurations only, therefore no results for $^{46}$Fe can be shown. 
The discrete isovector dipole strength distributions are smoothed by a Lorentzian of width $\Gamma=1$. Roughly speaking, the peak position of the
distributions from the two models differ by a few MeVs, which can be attributed to the different TRK enhancement factors. 
For $^{40, 48}$Ca, it is seen that the 
observed GDR peak lies between the predictions of the two models; 
CRPA+SLy4 describes the low energy region of the GDR fairly, while
 QRPA+D1S accounts for the tail of the GDR at high energies better.
In this sense both models are found equally appropriate in the case of the GDR. 
The fragmentation patterns are similar with both functionals. 

We stress that, athough the theoretical predictions for the GDR peak have been found correlated with the predicted enhancement factor, no correlation has been found between the pygmy dipole strength and the enhancement factor~\cite{Rei1999,ReN2010,ReN2013}. 

Next, we examine in detail the low-energy transitions with model QRPA+D1S and then we compare with CRPA results for further insights. 


\subsection{QRPA results} 

\begin{figure*}
\hspace*{-1.5cm}
\includegraphics[width=22cm,height=11cm]{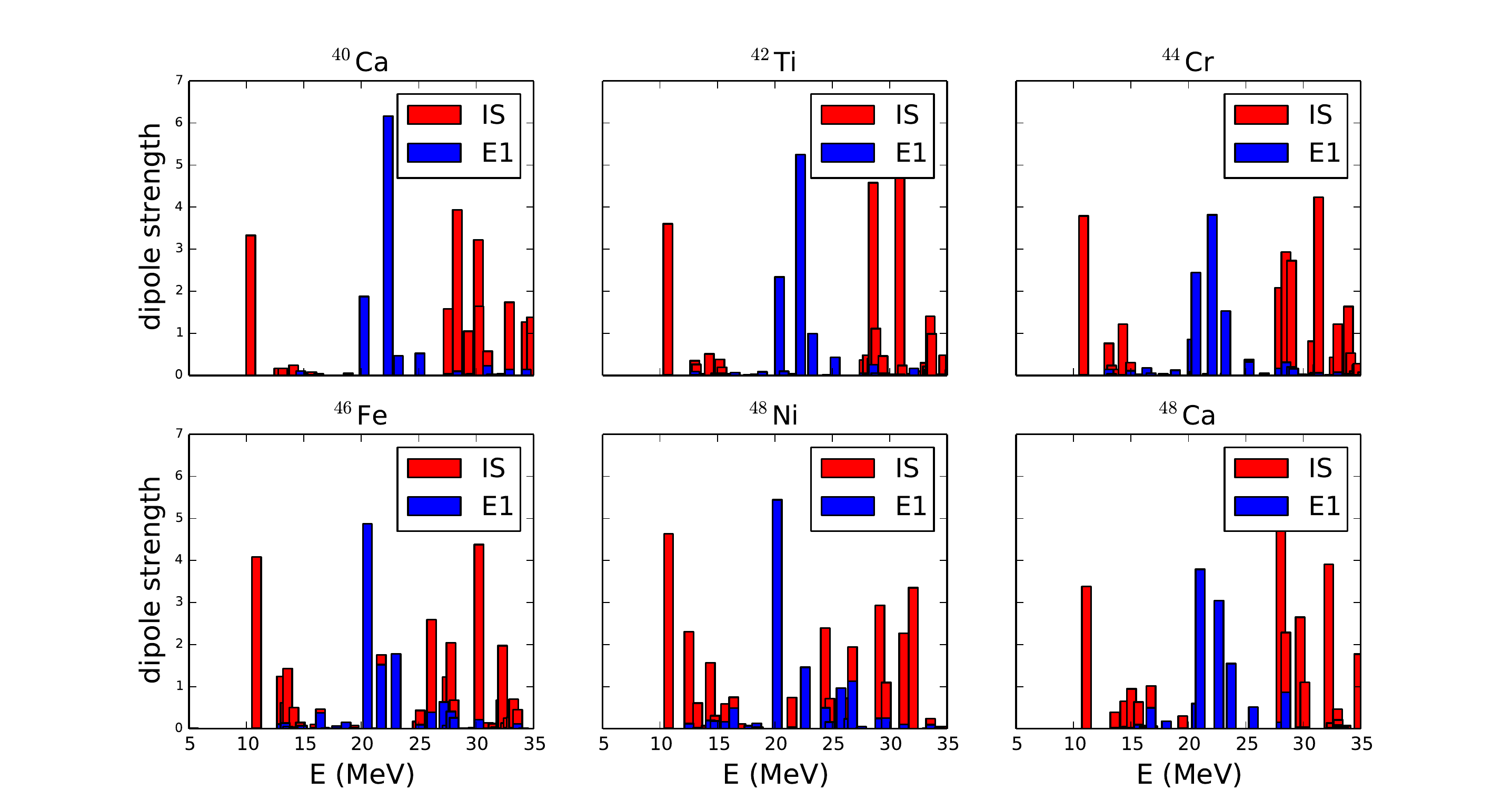}
\caption{(Color online)  QRPA IS and E1 dipole strength distributions in units of $e^2fm^6$ and $e^2fm^2$, respectively.
Note that the IS strength is divided by $250$ for presentation purposes only.}
\label{fig2}
\end{figure*}

\begin{figure}
\includegraphics[width=8cm,height=6cm]{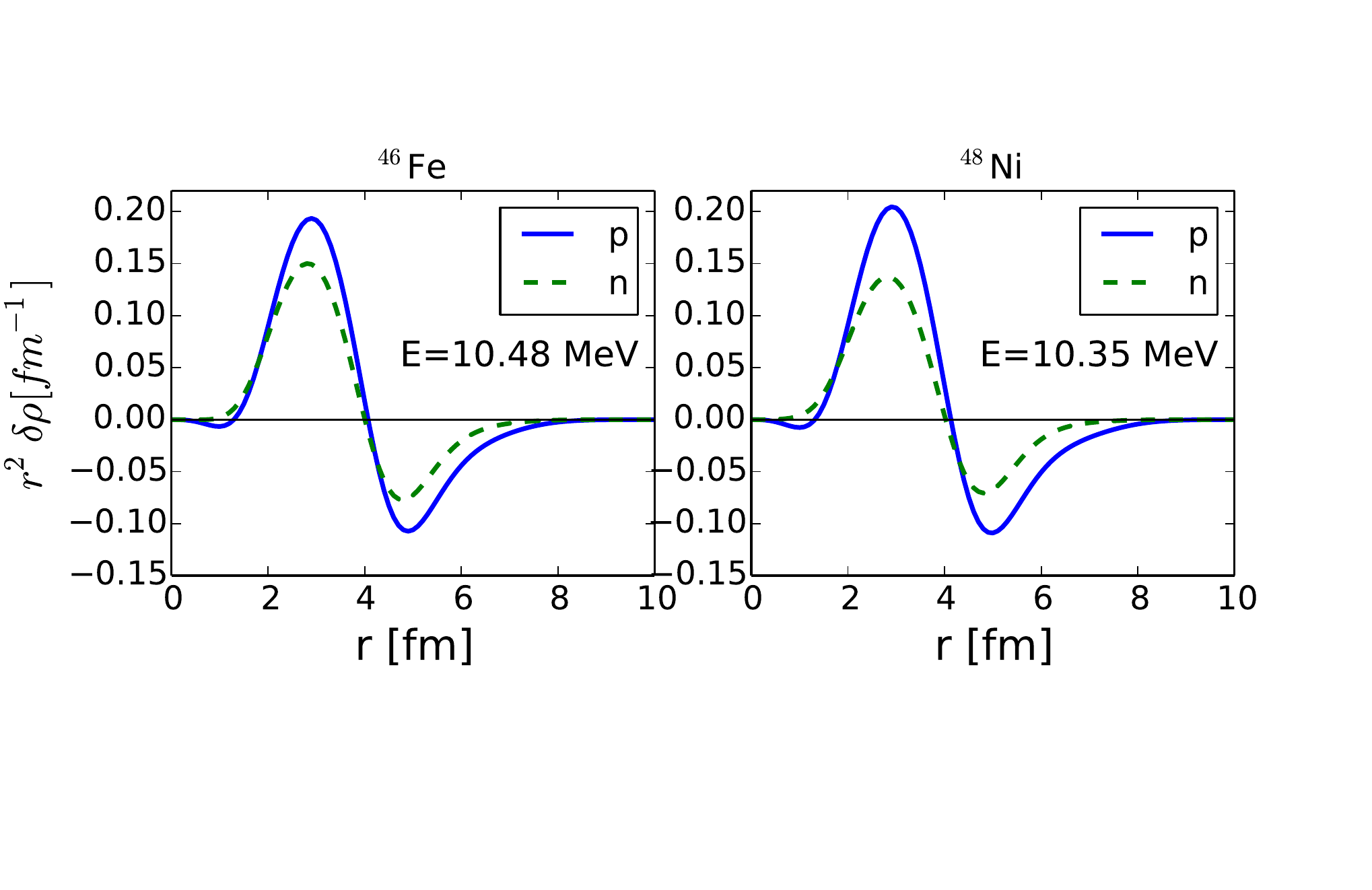}
\vspace*{-1.75cm}
\caption{(Color online)  Proton and neutron transition density of isoscalar low energy dipole states of $^{46}$Fe and
$^{48}$Ni at E=10.48 MeV and E=10.35 MeV, respectively.}
\label{fig3}
\end{figure}

Figure \ref{fig2} shows the IS and IV dipole transition strength from QRPA calculations for the $N=20$ isotones and for $^{48}$Ca. 
In the IV response the main feature in all nuclei is the GDR. 
In the IS response we observe at least two dominant features: The concentration of high-lying peaks, corresponding to the dipole compression mode~\cite{Hav2001}, 
and the strong transition  
around $10$ MeV ($9.9 ~{\rm MeV} < {\rm E} <10.8 ~{\rm MeV} $). 
The latter  
corresponds to the IS surface dipole mode, or IS low-energy dipole (IS-LED) mode, studied also in Refs.~\cite{PPR2011,PHP2012,PHP2014,Pap2015,PHR2015}. 
All structural properties of the IS-LED calculated in the present work are found in line with those studied in the above-cited publications, to which we refer the reader for details.
As an example, we show in Fig.~\ref{fig3} the transition densities of the IS-LED peaks  
in $^{46}$Fe and
$^{48}$Ni. 
The proton and neutron transition densities are of the IS-LED type, namely in phase and with a node near the surface. 
We have checked that the corresponding transition densities in the other nuclei examined here exhibit a
similar trend to the ones in Fig.~\ref{fig3}. 
In addition, the IS-LED peaks account for very small $E1$ strength as
manifest in Fig.~\ref{fig2} and Table \ref{table1}. 
The corresponding states have been observed in $^{40,48}$Ca around 7~MeV~\cite{Poe1992,Der2014} and other nuclei~\cite{Pap2015}. 
We point out that in $^{46}$Fe, $^{48}$Ni the IS-LED would be unbound because of the low particle emission threshold. 

\begin{table*}[ht]
\caption{Single-particle unit (single-proton for $N=20$ isotones and single neutron for
 $^{46,48}$Ca), $E1$ transition strength of the IS-LED, and $E1$ strengths summed  up to (13, 15, 17)
MeV, in units $e^2$fm$^2$ calculated with the QRPA+D1S model. 
The proton (or neutron) skin thickness is also given in fm. 
}
\begin{tabular}{r |c c c c c c }
\hline\hline
 & B$_{\rm s.p}$ &\, $B(E_1)$\,  & $\sum\limits_{E_i<13}  B(E_i) $\, &
 $\sum\limits_{E_i<15}  B(E_i) $\, & $\sum\limits_{E_i<17}  B(E_i) $  
& $|R_p - R_n|$ 
\\ [0.3ex]
\hline
          $^{40}$Ca & 0.188 & 0.004 & 0.008& 0.117 &0.179   & 0.042 
\\ 
          $^{42}$Ti & 0.177 & 0.008 & 0.116& 0.249 &0.361    & 0.106 
\\ 
          $^{44}$Cr & 0.166 & 0.009 & 0.189& 0.320 &0.576   & 0.159 
\\ [0.3ex] 
          $^{46}$Fe & 0.156 & 0.008 & 0.251& 0.421 &0.835   & 0.207 
\\ 
          $^{48}$Ni & 0.148 & 0.009 & 0.161& 0.553 &1.211   & 0.249 
\\ [0.3ex] 
\hline
          $^{46}$Ca & 0.156 & 0.003 & 0.050& 0.149 &0.443   & 0.108 
\\ 
          $^{48}$Ca & 0.148 & 0.006 & 0.006& 0.089 &0.763   & 0.145  
\\ [0.3ex] 
\hline
\end{tabular}
\label{table1}
\end{table*}

In Fig.~\ref{fig2} we observe also the effect of adding more and more protons to the symmetric nucleus $^{40}$Ca, namely increased transition strength, in both the IS and the IV channel, 
at energies between the IS-LED and the GDR. 
Nuclear properties that change as we add protons, are the asymmetry $|N-Z|/A$, the occupation number of the valence $\pi 0f_{7/2}$ orbital, 
and the proton separation energy. 
The effect of the latter can be isolated by a comparison of $^{48}$Ca and $^{48}$Ni.  
Indeed, if we compare the spectra of these two mirror nuclei  
and the values of low-energy strength 
tabulated in Table~\ref{table1}, we find that the barely bound $^{48}$Ni is much more strongly excited 
than the stable $^{48}$Ca. 
Similarly, if we compare 
the values of low-energy strength 
tabulated for  
the $A=46$ mirror nuclei, we find 
that $^{46}$Fe is much more strongly excited than the stable $^{46}$Ca at low energies. 
In the stable Ca isotopes, the $E1$ strength summed up to 15~MeV is comparable to the strength of a single-particle transition. 
 
\begin{figure}
\hspace*{-0.7cm}
\includegraphics[width=10cm,height=10cm]{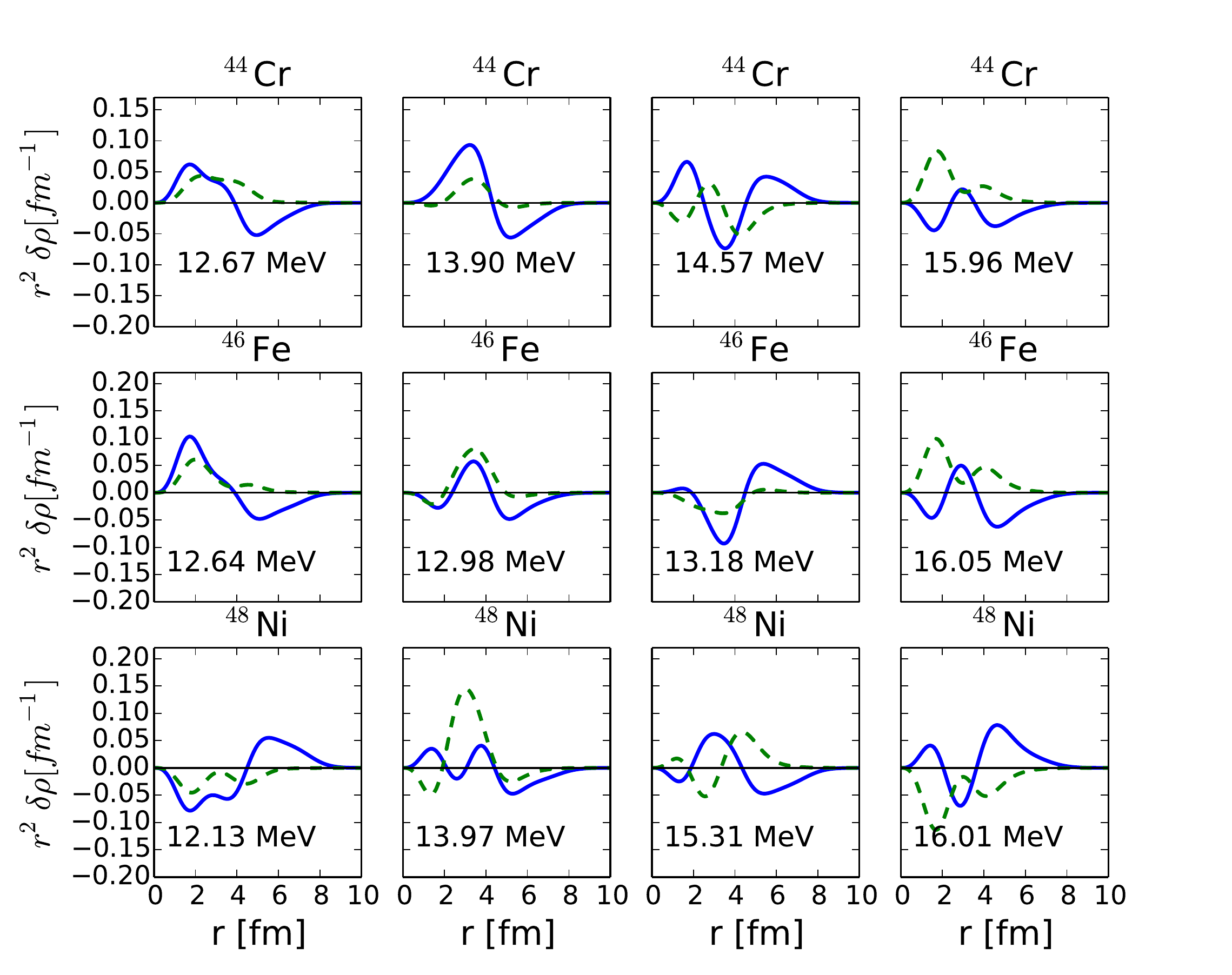}
\caption{(Color online)  Proton and neutron transition densities for selected isotones.}
\label{fig4}
\end{figure}

\begin{figure}
\hspace*{-0.7cm}
\includegraphics[width=10cm]{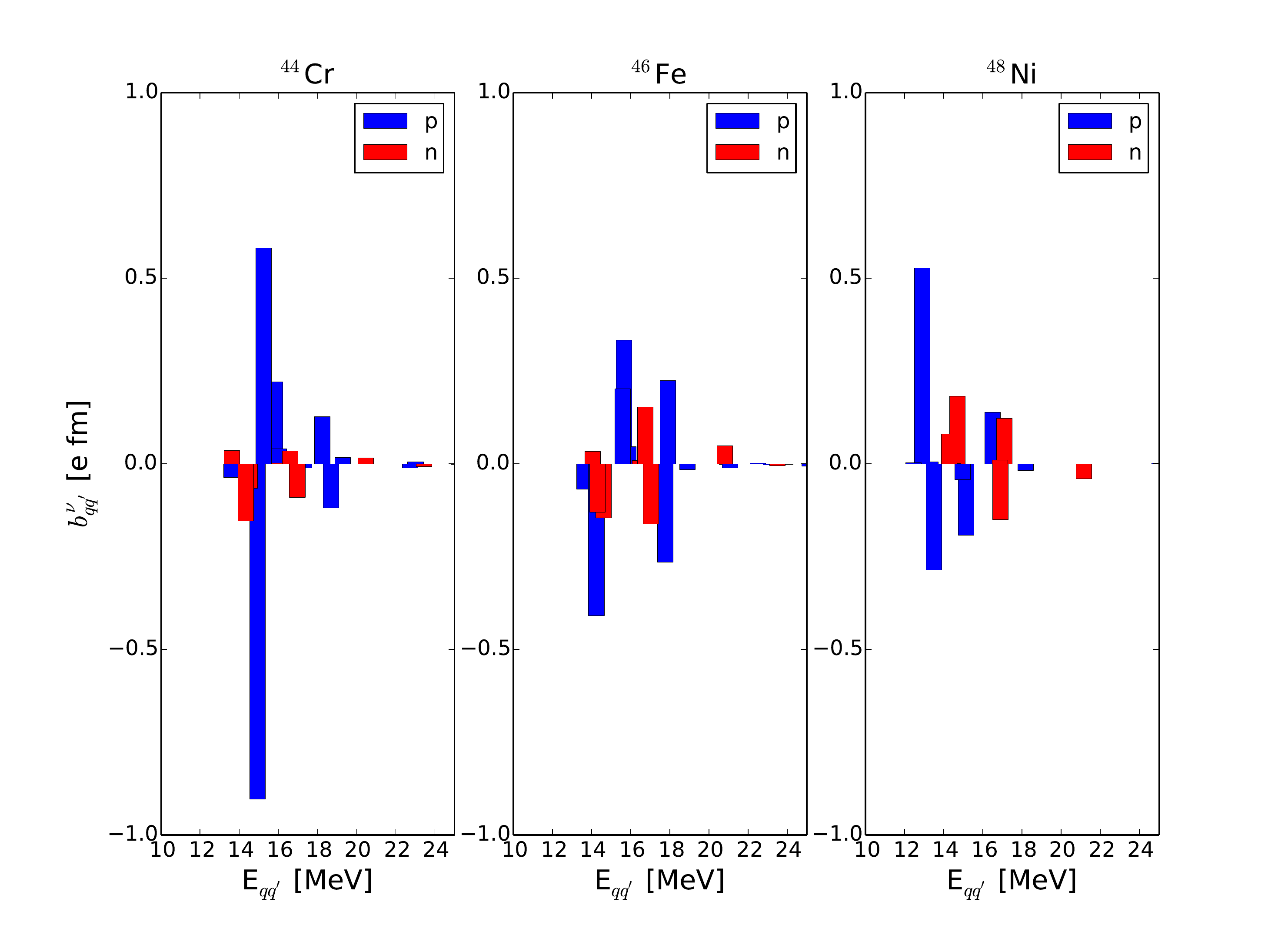}
\caption{(Color online) Two-quasiparticle configuration contributions to the transition matrix element of the pygmy states of
$^{44}$Cr, $^{46}$Fe, and $^{48}$Ni.}
\label{fig5}
\end{figure}

The fragmented resonant-like structure we have obtained between the IS-LED and the GDR in the proton-rich isotones represents a candidate for a proton-skin oscillation. 
Next, we assess this possibility. 
First we note that a skin oscillation would respond strongly in both channels, IS and IV. 
This is in analogy to the strong IS response expected of neutron-skin modes~\cite{End2010,VNP2012,LVL2014}. 
In Fig.~\ref{fig4} we display the transition densities of states which appear relatively strong in both channels, 
for the proton-rich isotones $^{44}$Cr, $^{46}$Fe, $^{48}$Ni. 
We have verified that the transition densities of omitted states do not show any skin-oscillation pattern or they resemble those of neighboring states shown in Fig.~\ref{fig4}.
The states beyond 15~MeV (rightmost states in Fig.~\ref{fig4}) show a strong IV character and can be considered as fragments of the GDR. 
Generally, the leftmost state for each isotone appears the most promising proton-pygmy candidate. 
In order to assess its collectivity we examine which quasiparticle configurations contribute to the $E1$ strength. 
In Fig.~\ref{fig5} we show the quantity $b_{qq'}^{\nu}$, Eq.~(\ref{Eq:bqqn}), for the IV operator $Q_{E1}$. 
We notice many contributions of destructive coherence in $^{46}$Fe, even if we consider only proton configurations, i.e., the uncorrected operator $Q_{E1}'$ (see below Eq.~(\ref{Eq:E1op})). 
In neutron-rich nuclei, neutron-skin oscillations have been associated with destructive coherence~\cite{Lan2009,VNP2012}. 
In $^{48}$Ni the dominant configuration is $\pi 0f_{7/2}\rightarrow \pi 1d_{5/2}$ ($\pi$ stands for protons). 
In $^{44}$Cr there is also a strong $\pi 0f_{7/2}\rightarrow \pi 0g_{9/2}$ component. 
In $^{46}$Fe, however, the transition is evidently not of single-particle character. 
The value in the third column of Table~\ref{table1} in comparison to the first column support the conclusion that the mode in question is stronger than a single-particle transition. 

We can conclude, based on our QRPA+D1S results, that we expect a proton pygmy mode to develop in proton-rich $N=20$ isotones, most unambiguously in $^{46}$Fe. 
Based on Figs.~\ref{fig2},~\ref{fig4} we expect a proton-pygmy resonance to appear somewhat fragmented and possibly surrounded by other $E1$ excitations. 

We note that the relativistic models of Refs.~\cite{PhysRevLett.94.182501,PPP2005} predict, as lowest-lying states, proton-pygmy modes of enhanced $E1$ strength. 
By contrast, the QRPA+D1S model predicts an ordinary IS-LED of very little $E1$ strength. 
The situation is reminiscent of that in neutron-rich nuclei, where neutron-skin vibrations are predicted for rather moderate neutron excess~\cite{PHP2014,PHR2015}. 
In proton-rich nuclei the two kinds of states may coexist in the continuum in reality. Next we address the treatment of the continuum.

\subsection{Comparisons with CRPA results} 

So far we have analyzed the QRPA eigenstates as dicrete and localized excitations. 
There are two effects missing from such a treatment, related to the fact that the particle-emission threshold is very low in the proton-rich isotones. 
First, practically all excited states are unbound and they may have a finite escape width. 
Second, protons can be excited to non-localized states near the Coulomb barrier (and neutrons to states of small positive energies), which cannot be efficiently described 
in a harmonic oscillator basis. 
The effect of the basis was already evident in the Hartree-Fock results of Fig.~\ref{fig0}.  
The CRPA can address the above-mentioned effects, albeit only for closed-shell configurations.
Next we examine our CRPA results on $^{48}$Ni and its mirror nucleus $^{48}$Ca in comparison to our QRPA results. 

\begin{figure*}
\hspace*{-0.7cm}
\includegraphics[width=16cm]{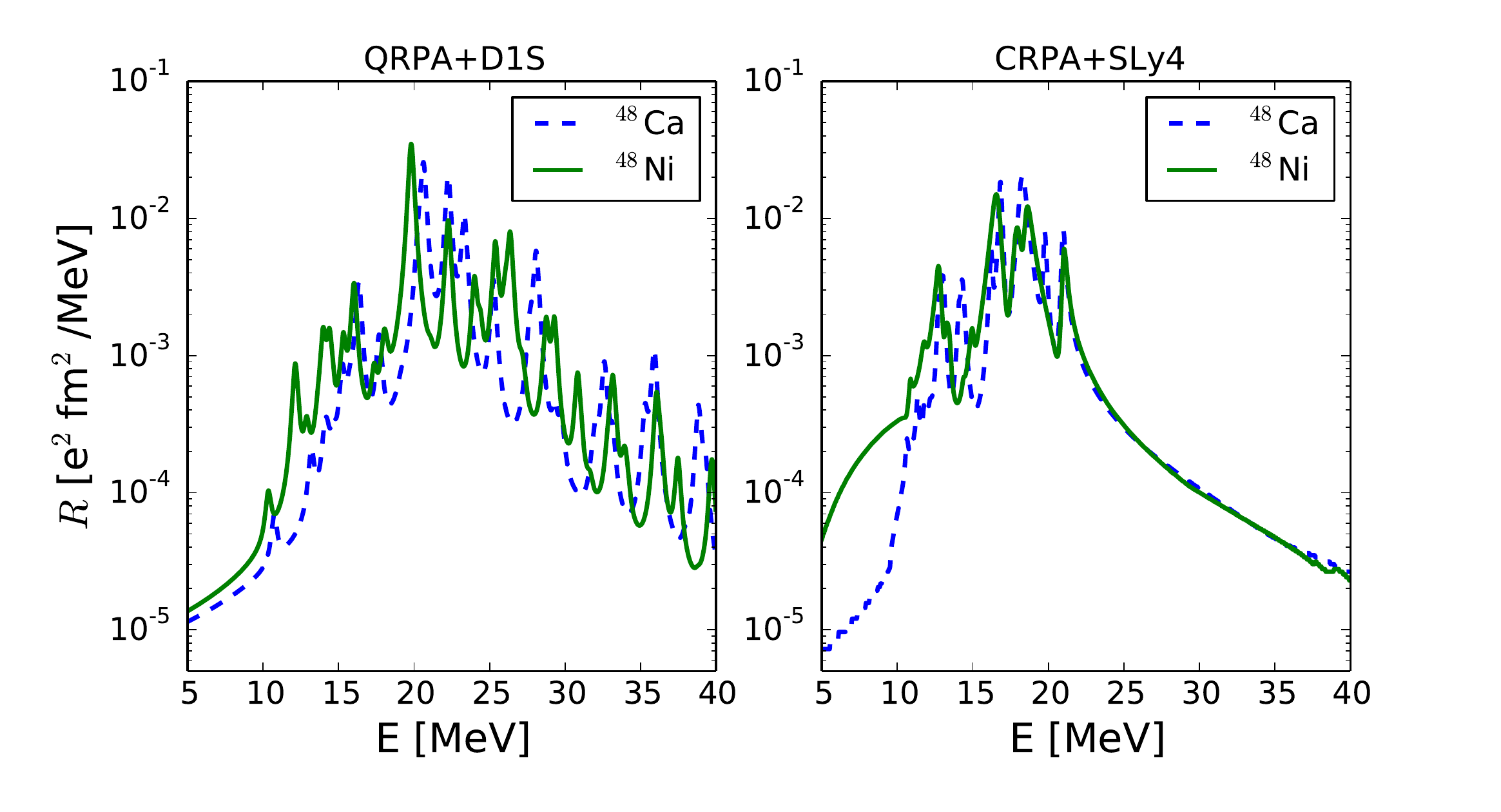}
\caption{(Color online) The $E1$ strength distributions for $^{48}$Ca and $^{48}$Ni in the two models.
A Lorentzian with $\Gamma=0.2$ is employed for the QRPA+D1S results. }
\label{fig6}
\end{figure*}

We begin by comparing the response of the two nuclei, in both models, in Fig.~\ref{fig6} and in Tables~\ref{table1},\ref{table2}. 
At this point we are discussing the top two rows of Table~\ref{table2} (SLy4 functional).
Each model predicts similar patterns for the GDR of the two nuclei, but enhanced low-energy $E1$ strength in $^{48}$Ni compared to $^{48}$Ca. 
In the case of CRPA, we recognize that the distribution shows a large escape width, as reported also in Ref.~\cite{PPP2005}.  

\begin{table}[ht]
\caption{$E1$ transition strength strength summed  up to (10, 13, 15)
MeV, in units $e^2$fm$^2$, and percentage points of the energy weighted sum below 10~MeV. 
}
\begin{tabular}{r |c c c c c}
\hline\hline
 &  $\int\limits_{E<10}\!\!\!  dB(E) $\, &  $\int\limits_{E<13}\!\!\!  dB(E) $\,&
  $\frac{100m_1^{\mathrm{low}}}{m_1^{\mathrm{tot}}}$ 
\\ [0.3ex]
\hline
SLy4:$^{48}$Ni &  0.242 & 1.241 &   1.01 
\\  
     $^{48}$Ca &  0.028 & 0.560 &   0.11 
\\ [0.3ex] 
BSk5:$^{48}$Ni &  0.254 & 0.950 &   1.08 
\\ 
     $^{48}$Ca &  0.029 & 0.528 &   0.12 
\\ [0.3ex] 
SkI3:$^{48}$Ni &  0.517 & 1.277 &   2.11 
\\ 
     $^{48}$Ca &  0.080 & 0.564 &   0.37 
\\ [0.3ex] 
\hline
\end{tabular}
\label{table2}
\end{table}

\begin{figure}
\hspace*{-0.7cm}
\includegraphics[width=9cm]{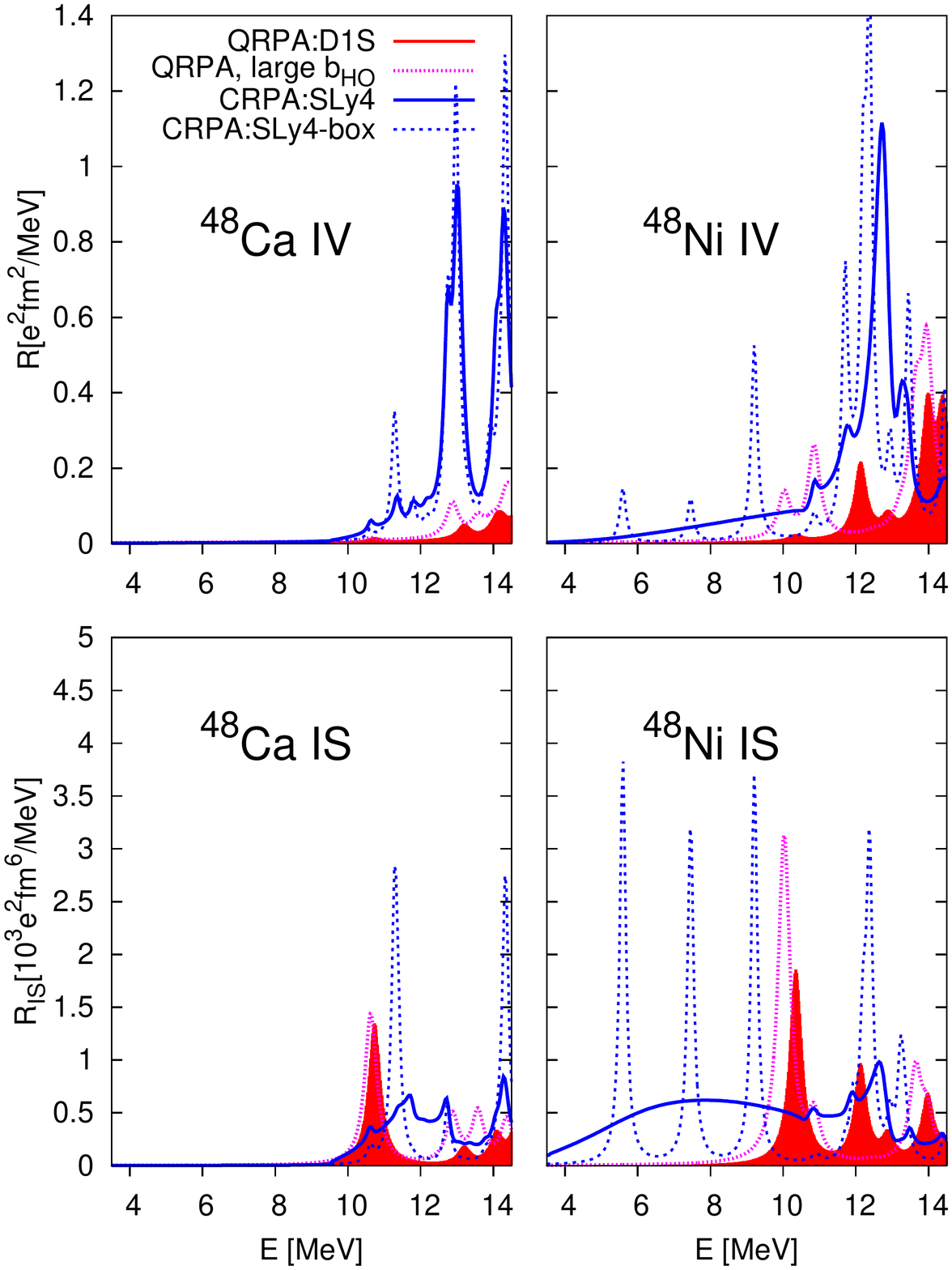}
\caption{(Color online) The low-energy $E1$ (or IV) and IS strength distributions for $^{48}$Ca and $^{48}$Ni in QRPA+D1S, QRPA+D1S with a larger length parameter $b_{\mathrm{HO}}$, the CRPA+SLy4, and the CRPA+SLy4 with box boundary conditions.
A Lorentzian with $\Gamma=0.2$ is employed. }
\label{fig7}
\end{figure}

In Fig.~\ref{fig7} we compare the results of the two models for each nucleus in the low-energy regime. 
There is a large amount of $E1$ strength predicted by CRPA in $^{48}$Ni below 10~MeV, corresponding to more than 1\% of the energy-weighted sum, but not in $^{48}$Ca. 
As evident from the values in Tables~\ref{table1},~\ref{table2}, 
this amount of strength is not described by QRPA+D1S, unless we take into account an energetic shift larger than the 3~MeV introduced in Sec.~\ref{Sec:Theory}. 


In Fig.~\ref{fig7} 
we include the results of the CRPA+SLy4 model with box boundary conditions imposed, to eliminate the escape width. 
For $^{48}$Ca the effect is very weak. 
For $^{48}$Ni the strongest effect is in the narrowing of the low-energy tail into three peaks, indicating that there are at least three configurations contributing to this 
broad structure, 
but the total strength does not change.   

The two models, namely D1S and SLy4, have similar symmetry properties (the SLy4 functional is even more asy-soft than the D1S) and the same nucleon effective mass. 
One might expect more strength predicted by the Gogny D1S functional, 
but we observe the opposite. Therefore, the only factor we can identify as enhancing the low-energy strength in the case of CRPA+SLy4 is the proper treatment of spatially extended configurations. 
%
The QRPA cannot account for the low-energy transition strength because of the bad convergence properties of the harmonic-oscillator basis for extended wavefunctions. 
Our statement is corroborated by a comparison of the QRPA results using different length parameters $b_{\mathrm{HO}}$.  
A larger value of $b_{\mathrm{HO}}$ (1.90~fm instead of 1.65~fm in $^{48}$Ni and 1.60~fm in $^{48}$Ca) 
has the effect of shifting pygmy strength to lower energy (and a rather weak overall effect at high energies) in $^{48}$Ni. 
The effect is weak but present in $^{48}$Ca. 
The sensitivity of the results to the length parameter signifies that 
a proper treatment of the threshold strength is very important in the description of pygmy proton strength.

\begin{figure}
\includegraphics[width=9cm]{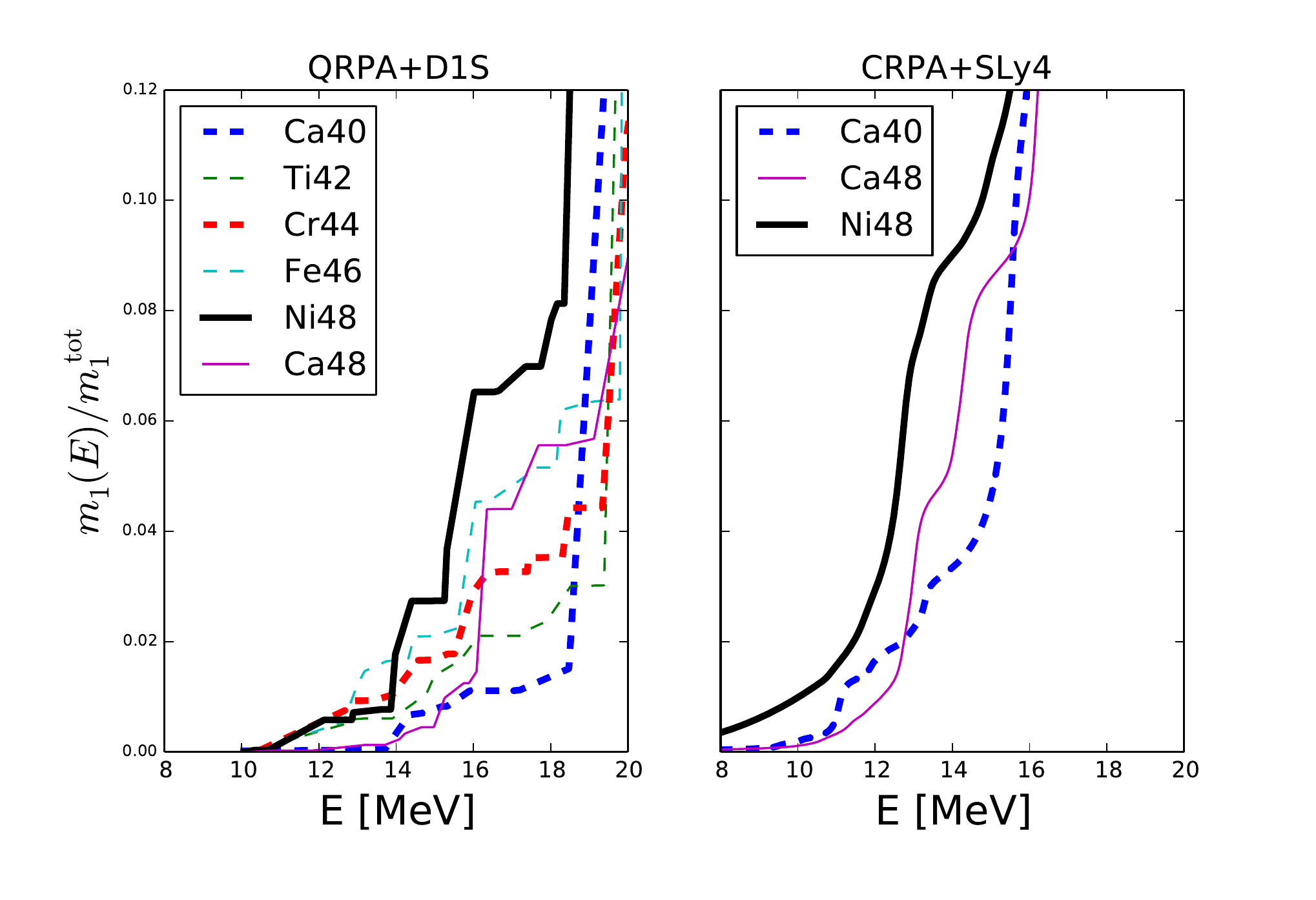}
\caption{(Color online) The energy-weighted sums of $N=20$ isotones and  $^{48}$Ca in the two approaches.}
\label{fig8}
\end{figure}

Finally, we discuss Fig.~\ref{fig8}, which displays the cummulative energy-weighted sum of $E1$ strength in both models, the total being normalized to one. According to the CCRPA+SLy4 model, 1\% of the energy-weighted sum can be expected below 10~MeV in $^{48}$Ni, in stark contrast to the mirror nucleus and to $^{40}$Ca. 
The importance of the low separation energy, rather than just the asymmetry, in generating this amount of strength is evident. 
According to the QRPA+D1S model, as long as we take into account an energetic shift of about 3~MeV, for the reasons discussed in Sec.~\ref{Sec:Theory}, a similar percentage of the $m_1^{\mathrm{tot}}$ is predicted in all isotones with $Z>20$, with the expection of $^{46}$Fe, where this amount may even double. (Without the 3~MeV shift the predicted amount of strength evidently vanishes.)
As we conclude from the values in Table~\ref{table1}, 1\% is consistent with one single-particle unit. 
$^{46}$Fe is therefore the most promising candidate for detecting a coherent skin mode. 
Nonetheless, a measurement in more-accessible lighter isotones would also be of value: If stronger transitions are detected, they might signify a collective effect. 
We note that our prediction for  $1\%$ of the EWS should be considered as a lower theoretical limit since it comes from asy-soft functionals with a low effective mass.

In the above we have focused on the isovector, $E1$ transition strenght. 
As is evident from Fig.~\ref{fig2} and our results with the Skyrme functional, Fig.~\ref{fig7}, 
the pygmy resonance and other proton transitions would be very strongly excited by isoscalar probes. 
An issue to keep in mind in that case is that a proton-skin mode and the ordinary IS-LED mode would both be unbound in proton-rich nuclei and likely indistinguishable. 
Helpful in this respect should be that the IS strength of the IS-LED transition in all isotones and in the mirror stable Ca isotopes should be approximately the same. 
Therefore, it can be considered approximately known, because the strength has been extracted in $^{40,48}$Ca~\cite{Poe1992,Der2014} and compares well with our calculations. 
A large excess of IS strength in proton-rich nuclei could then be attributed to exotic effects. 

\subsection{Other functionals} 

We have examined CRPA results with two more functionals and the results are tabulated in Table~\ref{table2}. 
The BSk5~\cite{GSB2003} functional has similar symmetry-energy properties as the D1S and SLy4 functionals~\cite{Dut2012}, 
but a higher nucleon effective mass, namely 0.92~$m$. 
The isovector effective mass is also somewhat higher for BSk5 (0.84$m$) compared to SLy4 (0.80$m$).  
The GDR is found somewhat shifted to lower energies in comparison to the SLy4 prediction, but almost the same $E1$ strength distribution is predicted below 10~MeV. 

In Ref.~\cite{PPP2005} it was observed that a Skyrme model of high isovector effective mass $m^{\ast}_v$ is required to obtain more strength and thus 
agreement with the relativistic models used in that work. 
The functional MSk7 with  $m^{\ast}_v=1.05m$ predicted the largest amount of low-energy strength, even though it is among the most asy-soft used there. 
We now test whether a combination of a low $m^{\ast}_v$ and a high $L$ can also give a large amount of strength. 
To this end we use SkI3, which predicts a low isoscalar effective mass 
but $L=100.53$ (the DD-ME1 and DD-ME2 relativistic models of Ref.~\cite{PPP2005} predict $L=55.42,\, 51.24$~MeV, respectively~\cite{Che2012}) 
and an isovector effective mass similar to the other models, namely $0.8m$. 
The results are given in Table~\ref{table2}. 
The amount of strength below 10~MeV is larger than twice the amount predicted by the other functionals. 
We therefore postulate, based on this exploratory calculation, that the asy-softness of the functional can affect the amount of pygmy dipole strength in an analogous way as the neutron-pygmy strength (see also Ref.~\cite{Bar2014}) and that  
both the isovector effective mass and the slope parameter can enhance the pygmy-dipole strength in proton rich nuclei. 
In Ref.~\cite{PPP2005} mainly the former is presumably at work in the case of the MSk7 functional, and the latter in the case of the relativistic models. 
Obviously, further systematic studies must be undertaken, along the lines of Refs.~\cite{INY2011,INY2013,ReN2010,Pie2012}, before the above trends can be established.

In beyond-mean-field models, an enhancement of the effective mass can be implicit through coupling of single-particle states to surface vibrations. 

\section{Summary and conclusion\label{Sec:Concl}}

We have studied theoretically the electric dipole response of the $N=20$ isotones at low energies (mainly below 10~MeV). 
We analyzed results of the QRPA solved in the harmonic oscillator basis and using the Gogny D1S interaction and of the CRPA solved in coordinate space using Skyrme functionals. 
These are widely used phenomenological methods, which in the future shall be complemented by more {\em ab initio} approaches~\cite{Bac2014}.

Larger amounts of $E1$ strength in the asymmetric $N=20$ isotones are predicted than the amounts of strength predicted or detected in equally asymmetric $Z=20$ mirror nuclei. The difference between the two mirror sets of asymmetric nuclei, $Z=20$ or $N=20$, is the much looser binding of the latter. Thus our results point unambiguously to the important role of structural effects, as opposed to global parameters like asymmetry, in determining the $E1$ spectrum at low energy.  

An exotic collective excitation is found most likely in $^{46}$Fe but perhaps also in 
$^{44}$Cr and $^{48}$Ni, exhausting more than 1\% of the energy-weighted summed strength. A similar amount of strength was predicted in Ref.~\cite{DaG2012} for the $N=20$ isotones using the relativistic continuum QRPA with a relatively soft relativistic model. 
In our above  prediction we take into account the results of the CRPA solved in coordinate space, the expectation that the energies of the low-lying states are overestimated by the QRPA model by about 3~MeV~\cite{PHP2012,PHR2015} and that the treatment of extended wavefunctions by the harmonic-oscillator basis in the same model is inefficient. 
A correct, converged treatment of threshold transitions (as in CRPA), and therefore of extended wavefunctions, is found important for the description of proton pygmy states since the proton emission treshold is extremely low. 

We consider the predicted amount of low-energy strength, namely $1\%$ of the EWS, as a lower theoretical limit, because it comes from asy-soft functionals with a low effective mass. 
We predict furthermore that the strength is likely distributed in a broad structure in the spectrum, which would respond very strongly to isoscalar probes. 

Based on an exploratory comparison of different functionals, we 
postulate, pending systematic studies, that the effective mass as well as the slope parameter $L$ can enhance the low-energy strength. 
A systematic and quantitative investigation of the two effects for proton-rich nuclei, along the lines of similar studies in neutron-rich nuclei~\cite{INY2011,INY2013,ReN2010,Pie2012} and with proper convergence, is highly desirable. 

\section*{Acknowledgments}
We thank H.Hergert for releasing the QRPA code. 
This work  was supported  by the Rare Isotope Science Project of Institute for Basic Science funded by Ministry of Science, ICT and Future Planning and
National Research Foundation of Korea (2013M7A1A1075764).
%

%

\begin{thebibliography}{43}%
\makeatletter
\providecommand \@ifxundefined [1]{%
 \@ifx{#1\undefined}
}%
\providecommand \@ifnum [1]{%
 \ifnum #1\expandafter \@firstoftwo
 \else \expandafter \@secondoftwo
 \fi
}%
\providecommand \@ifx [1]{%
 \ifx #1\expandafter \@firstoftwo
 \else \expandafter \@secondoftwo
 \fi
}%
\providecommand \natexlab [1]{#1}%
\providecommand \enquote  [1]{``#1''}%
\providecommand \bibnamefont  [1]{#1}%
\providecommand \bibfnamefont [1]{#1}%
\providecommand \citenamefont [1]{#1}%
\providecommand \href@noop [0]{\@secondoftwo}%
\providecommand \href [0]{\begingroup \@sanitize@url \@href}%
\providecommand \@href[1]{\@@startlink{#1}\@@href}%
\providecommand \@@href[1]{\endgroup#1\@@endlink}%
\providecommand \@sanitize@url [0]{\catcode `\\12\catcode `\$12\catcode
  `\&12\catcode `\#12\catcode `\^12\catcode `\_12\catcode `\%12\relax}%
\providecommand \@@startlink[1]{}%
\providecommand \@@endlink[0]{}%
\providecommand \url  [0]{\begingroup\@sanitize@url \@url }%
\providecommand \@url [1]{\endgroup\@href {#1}{\urlprefix }}%
\providecommand \urlprefix  [0]{URL }%
\providecommand \Eprint [0]{\href }%
\providecommand \doibase [0]{http://dx.doi.org/}%
\providecommand \selectlanguage [0]{\@gobble}%
\providecommand \bibinfo  [0]{\@secondoftwo}%
\providecommand \bibfield  [0]{\@secondoftwo}%
\providecommand \translation [1]{[#1]}%
\providecommand \BibitemOpen [0]{}%
\providecommand \bibitemStop [0]{}%
\providecommand \bibitemNoStop [0]{.\EOS\space}%
\providecommand \EOS [0]{\spacefactor3000\relax}%
\providecommand \BibitemShut  [1]{\csname bibitem#1\endcsname}%
\let\auto@bib@innerbib\@empty
\bibitem [{\citenamefont {Paar}\ \emph {et~al.}(2007)\citenamefont {Paar},
  \citenamefont {Vretenar}, \citenamefont {Khan},\ and\ \citenamefont
  {Col{\`o}}}]{PVK2007}%
  \BibitemOpen
  \bibfield  {author} {\bibinfo {author} {\bibfnamefont {N.}~\bibnamefont
  {Paar}}, \bibinfo {author} {\bibfnamefont {D.}~\bibnamefont {Vretenar}},
  \bibinfo {author} {\bibfnamefont {E.}~\bibnamefont {Khan}}, \ and\ \bibinfo
  {author} {\bibfnamefont {G.}~\bibnamefont {Col{\`o}}},\ }\href
  {http://stacks.iop.org/0034-4885/70/i=5/a=R02} {\bibfield  {journal}
  {\bibinfo  {journal} {Reports on Progress in Physics}\ }\textbf {\bibinfo
  {volume} {70}},\ \bibinfo {pages} {691} (\bibinfo {year} {2007})}\BibitemShut
  {NoStop}%
\bibitem [{\citenamefont {Krewald}\ and\ \citenamefont
  {Speth}(2009)}]{KrS2009}%
  \BibitemOpen
  \bibfield  {author} {\bibinfo {author} {\bibfnamefont {S.}~\bibnamefont
  {Krewald}}\ and\ \bibinfo {author} {\bibfnamefont {J.}~\bibnamefont
  {Speth}},\ }\href {\doibase 10.1142/S0218301309013750} {\bibfield  {journal}
  {\bibinfo  {journal} {International Journal of Modern Physics E}\ }\textbf
  {\bibinfo {volume} {18}},\ \bibinfo {pages} {1425} (\bibinfo {year}
  {2009})}\BibitemShut {NoStop}%
\bibitem [{\citenamefont {Savran}\ \emph {et~al.}(2013)\citenamefont {Savran},
  \citenamefont {Aumann},\ and\ \citenamefont {Zilges}}]{SAZ2013}%
  \BibitemOpen
  \bibfield  {author} {\bibinfo {author} {\bibfnamefont {D.}~\bibnamefont
  {Savran}}, \bibinfo {author} {\bibfnamefont {T.}~\bibnamefont {Aumann}}, \
  and\ \bibinfo {author} {\bibfnamefont {A.}~\bibnamefont {Zilges}},\ }\href
  {\doibase http://dx.doi.org/10.1016/j.ppnp.2013.02.003} {\bibfield  {journal}
  {\bibinfo  {journal} {Progress in Particle and Nuclear Physics}\ }\textbf
  {\bibinfo {volume} {70}},\ \bibinfo {pages} {210 } (\bibinfo {year}
  {2013})}\BibitemShut {NoStop}%
\bibitem [{\citenamefont {Paar}\ \emph {et~al.}(2005)\citenamefont {Paar},
  \citenamefont {Vretenar},\ and\ \citenamefont
  {Ring}}]{PhysRevLett.94.182501}%
  \BibitemOpen
  \bibfield  {author} {\bibinfo {author} {\bibfnamefont {N.}~\bibnamefont
  {Paar}}, \bibinfo {author} {\bibfnamefont {D.}~\bibnamefont {Vretenar}}, \
  and\ \bibinfo {author} {\bibfnamefont {P.}~\bibnamefont {Ring}},\ }\href
  {\doibase 10.1103/PhysRevLett.94.182501} {\bibfield  {journal} {\bibinfo
  {journal} {Phys. Rev. Lett.}\ }\textbf {\bibinfo {volume} {94}},\ \bibinfo
  {pages} {182501} (\bibinfo {year} {2005})}\BibitemShut {NoStop}%
\bibitem [{\citenamefont {Barbieri}\ \emph {et~al.}(2008)\citenamefont
  {Barbieri}, \citenamefont {Caurier}, \citenamefont {Langanke},\ and\
  \citenamefont {Mart\'inez-Pinedo}}]{PhysRevC.77.024304}%
  \BibitemOpen
  \bibfield  {author} {\bibinfo {author} {\bibfnamefont {C.}~\bibnamefont
  {Barbieri}}, \bibinfo {author} {\bibfnamefont {E.}~\bibnamefont {Caurier}},
  \bibinfo {author} {\bibfnamefont {K.}~\bibnamefont {Langanke}}, \ and\
  \bibinfo {author} {\bibfnamefont {G.}~\bibnamefont {Mart\'inez-Pinedo}},\
  }\href {\doibase 10.1103/PhysRevC.77.024304} {\bibfield  {journal} {\bibinfo
  {journal} {Phys. Rev. C}\ }\textbf {\bibinfo {volume} {77}},\ \bibinfo
  {pages} {024304} (\bibinfo {year} {2008})}\BibitemShut {NoStop}%
\bibitem [{\citenamefont {Paar}\ \emph {et~al.}(2004)\citenamefont {Paar},
  \citenamefont {Papakonstantinou}, \citenamefont {Ponomarev},\ and\
  \citenamefont {Wambach}}]{PPP2005}%
  \BibitemOpen
  \bibfield  {author} {\bibinfo {author} {\bibfnamefont {N.}~\bibnamefont
  {Paar}}, \bibinfo {author} {\bibfnamefont {P.}~\bibnamefont
  {Papakonstantinou}}, \bibinfo {author} {\bibfnamefont {V.}~\bibnamefont
  {Ponomarev}}, \ and\ \bibinfo {author} {\bibfnamefont {J.}~\bibnamefont
  {Wambach}},\ }\href@noop {} {\bibfield  {journal} {\bibinfo  {journal} {Phys.
  Lett.}\ }\textbf {\bibinfo {volume} {B624}},\ \bibinfo {pages} {195}
  (\bibinfo {year} {2004})}\BibitemShut {NoStop}%
\bibitem [{\citenamefont {Ma}\ \emph {et~al.}(2012)\citenamefont {Ma},
  \citenamefont {Dong}, \citenamefont {Yan}, \citenamefont {Zhang},\ and\
  \citenamefont {Zhang}}]{PhysRevC.85.044307}%
  \BibitemOpen
  \bibfield  {author} {\bibinfo {author} {\bibfnamefont {H.-L.}\ \bibnamefont
  {Ma}}, \bibinfo {author} {\bibfnamefont {B.-G.}\ \bibnamefont {Dong}},
  \bibinfo {author} {\bibfnamefont {Y.-L.}\ \bibnamefont {Yan}}, \bibinfo
  {author} {\bibfnamefont {H.-Q.}\ \bibnamefont {Zhang}}, \ and\ \bibinfo
  {author} {\bibfnamefont {X.-Z.}\ \bibnamefont {Zhang}},\ }\href {\doibase
  10.1103/PhysRevC.85.044307} {\bibfield  {journal} {\bibinfo  {journal} {Phys.
  Rev. C}\ }\textbf {\bibinfo {volume} {85}},\ \bibinfo {pages} {044307}
  (\bibinfo {year} {2012})}\BibitemShut {NoStop}%
\bibitem [{\citenamefont {Ma}\ and\ \citenamefont {Tian}(2011)}]{MaT2011}%
  \BibitemOpen
  \bibfield  {author} {\bibinfo {author} {\bibfnamefont {Z.}~\bibnamefont
  {Ma}}\ and\ \bibinfo {author} {\bibfnamefont {Y.}~\bibnamefont {Tian}},\
  }\href {\doibase 10.1007/s11433-011-4416-8} {\bibfield  {journal} {\bibinfo
  {journal} {Science China Physics, Mechanics and Astronomy}\ }\textbf
  {\bibinfo {volume} {54}},\ \bibinfo {pages} {49} (\bibinfo {year}
  {2011})}\BibitemShut {NoStop}%
\bibitem [{\citenamefont {Martini}\ \emph {et~al.}(2011)\citenamefont
  {Martini}, \citenamefont {P\'eru},\ and\ \citenamefont {Dupuis}}]{MPD2011}%
  \BibitemOpen
  \bibfield  {author} {\bibinfo {author} {\bibfnamefont {M.}~\bibnamefont
  {Martini}}, \bibinfo {author} {\bibfnamefont {S.}~\bibnamefont {P\'eru}}, \
  and\ \bibinfo {author} {\bibfnamefont {M.}~\bibnamefont {Dupuis}},\ }\href
  {\doibase 10.1103/PhysRevC.83.034309} {\bibfield  {journal} {\bibinfo
  {journal} {Phys. Rev. C}\ }\textbf {\bibinfo {volume} {83}},\ \bibinfo
  {pages} {034309} (\bibinfo {year} {2011})}\BibitemShut {NoStop}%
\bibitem [{\citenamefont {Daoutidis}\ and\ \citenamefont
  {Goriely}(2012)}]{DaG2012}%
  \BibitemOpen
  \bibfield  {author} {\bibinfo {author} {\bibfnamefont {I.}~\bibnamefont
  {Daoutidis}}\ and\ \bibinfo {author} {\bibfnamefont {S.}~\bibnamefont
  {Goriely}},\ }\href {\doibase 10.1103/PhysRevC.86.034328} {\bibfield
  {journal} {\bibinfo  {journal} {Phys. Rev. C}\ }\textbf {\bibinfo {volume}
  {86}},\ \bibinfo {pages} {034328} (\bibinfo {year} {2012})}\BibitemShut
  {NoStop}%
\bibitem [{\citenamefont {Lepyoshkina}(2013)}]{Lep2013}%
  \BibitemOpen
  \bibfield  {author} {\bibinfo {author} {\bibfnamefont {O.}~\bibnamefont
  {Lepyoshkina}},\ }\href@noop {} {\enquote {\bibinfo {title} {Coulomb
  excitation of proton-rich nuclei $^{32}${Ar} and $^{34}${Ar}},}\ } (\bibinfo
  {year} {2013}),\ \bibinfo {note} {{{P}}h.D. Thesis, T.U.Munich}\BibitemShut
  {NoStop}%
\bibitem [{\citenamefont {Papakonstantinou}\ \emph {et~al.}(2015)\citenamefont
  {Papakonstantinou}, \citenamefont {Hergert},\ and\ \citenamefont
  {Roth}}]{PHR2015}%
  \BibitemOpen
  \bibfield  {author} {\bibinfo {author} {\bibfnamefont {P.}~\bibnamefont
  {Papakonstantinou}}, \bibinfo {author} {\bibfnamefont {H.}~\bibnamefont
  {Hergert}}, \ and\ \bibinfo {author} {\bibfnamefont {R.}~\bibnamefont
  {Roth}},\ }\href@noop {} {\bibfield  {journal} {\bibinfo  {journal} {Phys.
  Rev.}\ }\textbf {\bibinfo {volume} {C92}},\ \bibinfo {pages} {034311}
  (\bibinfo {year} {2015})}\BibitemShut {NoStop}%
\bibitem [{\citenamefont {C{o'}}\ \emph {et~al.}(2013)\citenamefont {C{o'}},
  \citenamefont {De~Donno}, \citenamefont {Anguiano},\ and\ \citenamefont
  {Lallena}}]{CDA2013}%
  \BibitemOpen
  \bibfield  {author} {\bibinfo {author} {\bibfnamefont {G.}~\bibnamefont
  {C{o'}}}, \bibinfo {author} {\bibfnamefont {V.}~\bibnamefont {De~Donno}},
  \bibinfo {author} {\bibfnamefont {M.}~\bibnamefont {Anguiano}}, \ and\
  \bibinfo {author} {\bibfnamefont {A.}~\bibnamefont {Lallena}},\ }\href@noop
  {} {\bibfield  {journal} {\bibinfo  {journal} {Phys. Rev. C}\ }\textbf
  {\bibinfo {volume} {87}},\ \bibinfo {pages} {034305} (\bibinfo {year}
  {2013})}\BibitemShut {NoStop}%
\bibitem [{\citenamefont {Berger}\ \emph {et~al.}(1991)\citenamefont {Berger},
  \citenamefont {Girod},\ and\ \citenamefont {Gogny}}]{BGG1991}%
  \BibitemOpen
  \bibfield  {author} {\bibinfo {author} {\bibfnamefont {J.}~\bibnamefont
  {Berger}}, \bibinfo {author} {\bibfnamefont {M.}~\bibnamefont {Girod}}, \
  and\ \bibinfo {author} {\bibfnamefont {D.}~\bibnamefont {Gogny}},\
  }\href@noop {} {\bibfield  {journal} {\bibinfo  {journal} {Comp. Phys.
  Comm.}\ }\textbf {\bibinfo {volume} {63}},\ \bibinfo {pages} {365} (\bibinfo
  {year} {1991})}\BibitemShut {NoStop}%
\bibitem [{\citenamefont {Papakonstantinou}\ \emph {et~al.}(2011)\citenamefont
  {Papakonstantinou}, \citenamefont {Ponomarev}, \citenamefont {Roth},\ and\
  \citenamefont {Wambach}}]{PPR2011}%
  \BibitemOpen
  \bibfield  {author} {\bibinfo {author} {\bibfnamefont {P.}~\bibnamefont
  {Papakonstantinou}}, \bibinfo {author} {\bibfnamefont {V.}~\bibnamefont
  {Ponomarev}}, \bibinfo {author} {\bibfnamefont {R.}~\bibnamefont {Roth}}, \
  and\ \bibinfo {author} {\bibfnamefont {J.}~\bibnamefont {Wambach}},\
  }\href@noop {} {\bibfield  {journal} {\bibinfo  {journal} {Eur. Phys. J.}\
  }\textbf {\bibinfo {volume} {A47}},\ \bibinfo {pages} {14} (\bibinfo {year}
  {2011})}\BibitemShut {NoStop}%
\bibitem [{\citenamefont {Papakonstantinou}\ \emph {et~al.}(2012)\citenamefont
  {Papakonstantinou}, \citenamefont {Hergert}, \citenamefont {Ponomarev},\ and\
  \citenamefont {Roth}}]{PHP2012}%
  \BibitemOpen
  \bibfield  {author} {\bibinfo {author} {\bibfnamefont {P.}~\bibnamefont
  {Papakonstantinou}}, \bibinfo {author} {\bibfnamefont {H.}~\bibnamefont
  {Hergert}}, \bibinfo {author} {\bibfnamefont {V.}~\bibnamefont {Ponomarev}},
  \ and\ \bibinfo {author} {\bibfnamefont {R.}~\bibnamefont {Roth}},\
  }\href@noop {} {\bibfield  {journal} {\bibinfo  {journal} {Phys. Lett. B}\
  }\textbf {\bibinfo {volume} {709}},\ \bibinfo {pages} {270} (\bibinfo {year}
  {2012})}\BibitemShut {NoStop}%
\bibitem [{\citenamefont {Papakonstantinou}\ \emph {et~al.}(2014)\citenamefont
  {Papakonstantinou}, \citenamefont {Hergert}, \citenamefont {Ponomarev},\ and\
  \citenamefont {Roth}}]{PHP2014}%
  \BibitemOpen
  \bibfield  {author} {\bibinfo {author} {\bibfnamefont {P.}~\bibnamefont
  {Papakonstantinou}}, \bibinfo {author} {\bibfnamefont {H.}~\bibnamefont
  {Hergert}}, \bibinfo {author} {\bibfnamefont {V.}~\bibnamefont {Ponomarev}},
  \ and\ \bibinfo {author} {\bibfnamefont {R.}~\bibnamefont {Roth}},\
  }\href@noop {} {\bibfield  {journal} {\bibinfo  {journal} {Phys. Rev.}\
  }\textbf {\bibinfo {volume} {C89}},\ \bibinfo {pages} {034306} (\bibinfo
  {year} {2014})}\BibitemShut {NoStop}%
\bibitem [{\citenamefont {Derya}\ \emph {et~al.}(2014)\citenamefont {Derya}
  \emph {et~al.}}]{Der2014}%
  \BibitemOpen
  \bibfield  {author} {\bibinfo {author} {\bibfnamefont {V.}~\bibnamefont
  {Derya}} \emph {et~al.},\ }\href@noop {} {\bibfield  {journal} {\bibinfo
  {journal} {Phys. Lett. B}\ }\textbf {\bibinfo {volume} {730}},\ \bibinfo
  {pages} {288} (\bibinfo {year} {2014})}\BibitemShut {NoStop}%
\bibitem [{\citenamefont {Reinhard}(1991)}]{ReXX}%
  \BibitemOpen
  \bibfield  {author} {\bibinfo {author} {\bibfnamefont {P.-G.}\ \bibnamefont
  {Reinhard}},\ }\href@noop {} {\bibfield  {journal} {\bibinfo  {journal} {{in
  Computational Nuclear Physics I - Nuclear Structure}}\ } (\bibinfo {year}
  {ed. K. Langanke, J.E. Maruhn and S.E. Koonin (Springer, New York 1991)})},\
  \bibinfo {note} {p.28}\BibitemShut {NoStop}%
\bibitem [{\citenamefont {Chen}\ \emph {et~al.}(2012)\citenamefont {Chen},
  \citenamefont {Cai}, \citenamefont {Chen}, \citenamefont {Li}, \citenamefont
  {Li},\ and\ \citenamefont {Xu}}]{Che2012}%
  \BibitemOpen
  \bibfield  {author} {\bibinfo {author} {\bibfnamefont {R.}~\bibnamefont
  {Chen}}, \bibinfo {author} {\bibfnamefont {B.-J.}\ \bibnamefont {Cai}},
  \bibinfo {author} {\bibfnamefont {L.-W.}\ \bibnamefont {Chen}}, \bibinfo
  {author} {\bibfnamefont {B.-A.}\ \bibnamefont {Li}}, \bibinfo {author}
  {\bibfnamefont {X.-H.}\ \bibnamefont {Li}}, \ and\ \bibinfo {author}
  {\bibfnamefont {C.}~\bibnamefont {Xu}},\ }\href {\doibase
  10.1103/PhysRevC.85.024305} {\bibfield  {journal} {\bibinfo  {journal} {Phys.
  Rev. C}\ }\textbf {\bibinfo {volume} {85}},\ \bibinfo {pages} {024305}
  (\bibinfo {year} {2012})}\BibitemShut {NoStop}%
\bibitem [{\citenamefont {P{\'e}ru}\ \emph {et~al.}(2005)\citenamefont
  {P{\'e}ru}, \citenamefont {Berger},\ and\ \citenamefont
  {Bortignon}}]{PBB2005}%
  \BibitemOpen
  \bibfield  {author} {\bibinfo {author} {\bibfnamefont {S.}~\bibnamefont
  {P{\'e}ru}}, \bibinfo {author} {\bibfnamefont {J.}~\bibnamefont {Berger}}, \
  and\ \bibinfo {author} {\bibfnamefont {P.}~\bibnamefont {Bortignon}},\
  }\href@noop {} {\bibfield  {journal} {\bibinfo  {journal} {Eur. Phys. J.}\
  }\textbf {\bibinfo {volume} {A26}},\ \bibinfo {pages} {25} (\bibinfo {year}
  {2005})}\BibitemShut {NoStop}%
\bibitem [{\citenamefont {Reinhard}(1999)}]{Rei1999}%
  \BibitemOpen
  \bibfield  {author} {\bibinfo {author} {\bibfnamefont {P.-G.}\ \bibnamefont
  {Reinhard}},\ }\href@noop {} {\bibfield  {journal} {\bibinfo  {journal}
  {Nucl. Phys. {\bf A649}}\ } (\bibinfo {year} {(1999)})},\ \bibinfo {note}
  {305c}\BibitemShut {NoStop}%
\bibitem [{\citenamefont {Reinhard}\ and\ \citenamefont
  {Nazarewicz}(2010)}]{ReN2010}%
  \BibitemOpen
  \bibfield  {author} {\bibinfo {author} {\bibfnamefont {P.}~\bibnamefont
  {Reinhard}}\ and\ \bibinfo {author} {\bibfnamefont {W.}~\bibnamefont
  {Nazarewicz}},\ }\href@noop {} {\bibfield  {journal} {\bibinfo  {journal}
  {Phys. Rev. C}\ }\textbf {\bibinfo {volume} {81}},\ \bibinfo {pages}
  {051303(R)} (\bibinfo {year} {2010})}\BibitemShut {NoStop}%
\bibitem [{\citenamefont {Reinhard}\ and\ \citenamefont
  {Nazarewicz}(2013)}]{ReN2013}%
  \BibitemOpen
  \bibfield  {author} {\bibinfo {author} {\bibfnamefont {P.}~\bibnamefont
  {Reinhard}}\ and\ \bibinfo {author} {\bibfnamefont {W.}~\bibnamefont
  {Nazarewicz}},\ }\href@noop {} {\bibfield  {journal} {\bibinfo  {journal}
  {Phys. Rev. C}\ }\textbf {\bibinfo {volume} {87}},\ \bibinfo {pages} {014324}
  (\bibinfo {year} {2013})}\BibitemShut {NoStop}%
\bibitem [{\citenamefont {Hergert}\ \emph {et~al.}(2011)\citenamefont
  {Hergert}, \citenamefont {Papakonstantinou},\ and\ \citenamefont
  {Roth}}]{HPR2011}%
  \BibitemOpen
  \bibfield  {author} {\bibinfo {author} {\bibfnamefont {H.}~\bibnamefont
  {Hergert}}, \bibinfo {author} {\bibfnamefont {P.}~\bibnamefont
  {Papakonstantinou}}, \ and\ \bibinfo {author} {\bibfnamefont
  {R.}~\bibnamefont {Roth}},\ }\href@noop {} {\bibfield  {journal} {\bibinfo
  {journal} {Phys. Rev. C}\ }\textbf {\bibinfo {volume} {83}},\ \bibinfo
  {pages} {064317} (\bibinfo {year} {2011})}\BibitemShut {NoStop}%
\bibitem [{\citenamefont {Goriely}\ and\ \citenamefont {Khan}(2002)}]{GoK2002}%
  \BibitemOpen
  \bibfield  {author} {\bibinfo {author} {\bibfnamefont {S.}~\bibnamefont
  {Goriely}}\ and\ \bibinfo {author} {\bibfnamefont {E.}~\bibnamefont {Khan}},\
  }\href@noop {} {\bibfield  {journal} {\bibinfo  {journal} {Nucl. Phys.}\
  }\textbf {\bibinfo {volume} {A706}},\ \bibinfo {pages} {217} (\bibinfo {year}
  {2002})}\BibitemShut {NoStop}%
\bibitem [{\citenamefont {Bertsch}\ and\ \citenamefont {Tsai}(1974)}]{BeT1974}%
  \BibitemOpen
  \bibfield  {author} {\bibinfo {author} {\bibfnamefont {G.}~\bibnamefont
  {Bertsch}}\ and\ \bibinfo {author} {\bibfnamefont {S.}~\bibnamefont {Tsai}},\
  }\href@noop {} {\bibfield  {journal} {\bibinfo  {journal} {Phys. Lett.}\
  }\textbf {\bibinfo {volume} {50B}},\ \bibinfo {pages} {319} (\bibinfo {year}
  {1974})}\BibitemShut {NoStop}%
\bibitem [{\citenamefont {Wang}\ \emph {et~al.}(2012)\citenamefont {Wang},
  \citenamefont {Audi}, \citenamefont {Wapstra}, \citenamefont {Kondev},
  \citenamefont {MacCormick}, \citenamefont {Xu},\ and\ \citenamefont
  {Pfeiffer}}]{AME2012}%
  \BibitemOpen
  \bibfield  {author} {\bibinfo {author} {\bibfnamefont {M.}~\bibnamefont
  {Wang}}, \bibinfo {author} {\bibfnamefont {G.}~\bibnamefont {Audi}}, \bibinfo
  {author} {\bibfnamefont {A.}~\bibnamefont {Wapstra}}, \bibinfo {author}
  {\bibfnamefont {F.}~\bibnamefont {Kondev}}, \bibinfo {author} {\bibfnamefont
  {M.}~\bibnamefont {MacCormick}}, \bibinfo {author} {\bibfnamefont
  {X.}~\bibnamefont {Xu}}, \ and\ \bibinfo {author} {\bibfnamefont
  {B.}~\bibnamefont {Pfeiffer}},\ }\href
  {http://stacks.iop.org/1674-1137/36/i=12/a=003} {\bibfield  {journal}
  {\bibinfo  {journal} {Chinese Physics C}\ }\textbf {\bibinfo {volume} {36}},\
  \bibinfo {pages} {1603} (\bibinfo {year} {2012})}\BibitemShut {NoStop}%
\bibitem [{CDF()}]{CDFE}%
  \BibitemOpen
  \href@noop {} {}\bibinfo {note} {CDFE database,
  http://cdfe.sinp.msu.ru/services/gdrsearch.html}\BibitemShut {NoStop}%
\bibitem [{\citenamefont {Harakeh}\ and\ \citenamefont {van~der
  Woude}(2001)}]{Hav2001}%
  \BibitemOpen
  \bibfield  {author} {\bibinfo {author} {\bibfnamefont {M.}~\bibnamefont
  {Harakeh}}\ and\ \bibinfo {author} {\bibfnamefont {A.}~\bibnamefont {van~der
  Woude}},\ }\href@noop {} {\emph {\bibinfo {title} {Giant Resonances}}}\
  (\bibinfo  {publisher} {Oxford Science Publications},\ \bibinfo {year}
  {2001})\BibitemShut {NoStop}%
\bibitem [{\citenamefont {Papakonstantinou}(2015)}]{Pap2015}%
  \BibitemOpen
  \bibfield  {author} {\bibinfo {author} {\bibfnamefont {P.}~\bibnamefont
  {Papakonstantinou}},\ }\href@noop {} {\bibfield  {journal} {\bibinfo
  {journal} {JPS Conf. Proc.}\ }\textbf {\bibinfo {volume} {6}},\ \bibinfo
  {pages} {030094} (\bibinfo {year} {2015})}\BibitemShut {NoStop}%
\bibitem [{\citenamefont {Poelhekken}\ \emph {et~al.}(1992)\citenamefont
  {Poelhekken} \emph {et~al.}}]{Poe1992}%
  \BibitemOpen
  \bibfield  {author} {\bibinfo {author} {\bibfnamefont {T.}~\bibnamefont
  {Poelhekken}} \emph {et~al.},\ }\href@noop {} {\bibfield  {journal} {\bibinfo
   {journal} {Phys. Lett.}\ }\textbf {\bibinfo {volume} {B278}},\ \bibinfo
  {pages} {423} (\bibinfo {year} {1992})}\BibitemShut {NoStop}%
\bibitem [{\citenamefont {Endres}\ \emph {et~al.}(2010)\citenamefont {Endres}
  \emph {et~al.}}]{End2010}%
  \BibitemOpen
  \bibfield  {author} {\bibinfo {author} {\bibfnamefont {J.}~\bibnamefont
  {Endres}} \emph {et~al.},\ }\href@noop {} {\bibfield  {journal} {\bibinfo
  {journal} {Phys. Rev. Lett.}\ }\textbf {\bibinfo {volume} {105}},\ \bibinfo
  {pages} {212503} (\bibinfo {year} {2010})}\BibitemShut {NoStop}%
\bibitem [{\citenamefont {Vretenar}\ \emph {et~al.}(2012)\citenamefont
  {Vretenar}, \citenamefont {Niu}, \citenamefont {Paar},\ and\ \citenamefont
  {Meng}}]{VNP2012}%
  \BibitemOpen
  \bibfield  {author} {\bibinfo {author} {\bibfnamefont {D.}~\bibnamefont
  {Vretenar}}, \bibinfo {author} {\bibfnamefont {Y.}~\bibnamefont {Niu}},
  \bibinfo {author} {\bibfnamefont {N.}~\bibnamefont {Paar}}, \ and\ \bibinfo
  {author} {\bibfnamefont {J.}~\bibnamefont {Meng}},\ }\href@noop {} {\bibfield
   {journal} {\bibinfo  {journal} {Phys. Rev. C}\ }\textbf {\bibinfo {volume}
  {85}},\ \bibinfo {pages} {044317} (\bibinfo {year} {2012})}\BibitemShut
  {NoStop}%
\bibitem [{\citenamefont {Lanza}\ \emph {et~al.}(2014)\citenamefont {Lanza},
  \citenamefont {Vitturi}, \citenamefont {Litvinova},\ and\ \citenamefont
  {Savran}}]{LVL2014}%
  \BibitemOpen
  \bibfield  {author} {\bibinfo {author} {\bibfnamefont {E.}~\bibnamefont
  {Lanza}}, \bibinfo {author} {\bibfnamefont {A.}~\bibnamefont {Vitturi}},
  \bibinfo {author} {\bibfnamefont {E.}~\bibnamefont {Litvinova}}, \ and\
  \bibinfo {author} {\bibfnamefont {D.}~\bibnamefont {Savran}},\ }\href@noop {}
  {\bibfield  {journal} {\bibinfo  {journal} {Phys. Rev.}\ }\textbf {\bibinfo
  {volume} {C89}},\ \bibinfo {pages} {041601(R)} (\bibinfo {year}
  {2014})}\BibitemShut {NoStop}%
\bibitem [{\citenamefont {Lanza}\ \emph {et~al.}(2009)\citenamefont {Lanza},
  \citenamefont {Catara}, \citenamefont {Gambacurta}, \citenamefont
  {Andr{\'e}},\ and\ \citenamefont {Chomaz}}]{Lan2009}%
  \BibitemOpen
  \bibfield  {author} {\bibinfo {author} {\bibfnamefont {E.}~\bibnamefont
  {Lanza}}, \bibinfo {author} {\bibfnamefont {F.}~\bibnamefont {Catara}},
  \bibinfo {author} {\bibfnamefont {D.}~\bibnamefont {Gambacurta}}, \bibinfo
  {author} {\bibfnamefont {M.}~\bibnamefont {Andr{\'e}}}, \ and\ \bibinfo
  {author} {\bibfnamefont {P.}~\bibnamefont {Chomaz}},\ }\href@noop {}
  {\bibfield  {journal} {\bibinfo  {journal} {Phys. Rev. C}\ }\textbf {\bibinfo
  {volume} {79}},\ \bibinfo {pages} {054615} (\bibinfo {year}
  {2009})}\BibitemShut {NoStop}%
\bibitem [{\citenamefont {Goriely}\ \emph {et~al.}(2003)\citenamefont
  {Goriely}, \citenamefont {Samyn}, \citenamefont {Bender},\ and\ \citenamefont
  {Pearson}}]{GSB2003}%
  \BibitemOpen
  \bibfield  {author} {\bibinfo {author} {\bibfnamefont {S.}~\bibnamefont
  {Goriely}}, \bibinfo {author} {\bibfnamefont {M.}~\bibnamefont {Samyn}},
  \bibinfo {author} {\bibfnamefont {M.}~\bibnamefont {Bender}}, \ and\ \bibinfo
  {author} {\bibfnamefont {J.}~\bibnamefont {Pearson}},\ }\href@noop {}
  {\bibfield  {journal} {\bibinfo  {journal} {Phys. Rev.}\ }\textbf {\bibinfo
  {volume} {C68}},\ \bibinfo {pages} {054325} (\bibinfo {year}
  {2003})}\BibitemShut {NoStop}%
\bibitem [{\citenamefont {Dutra}\ \emph {et~al.}(2012)\citenamefont {Dutra},
  \citenamefont {Louren\ifmmode~\mbox{\c{c}}\else \c{c}\fi{}o}, \citenamefont
  {S\'a~Martins}, \citenamefont {Delfino}, \citenamefont {Stone},\ and\
  \citenamefont {Stevenson}}]{Dut2012}%
  \BibitemOpen
  \bibfield  {author} {\bibinfo {author} {\bibfnamefont {M.}~\bibnamefont
  {Dutra}}, \bibinfo {author} {\bibfnamefont {O.}~\bibnamefont
  {Louren\ifmmode~\mbox{\c{c}}\else \c{c}\fi{}o}}, \bibinfo {author}
  {\bibfnamefont {J.~S.}\ \bibnamefont {S\'a~Martins}}, \bibinfo {author}
  {\bibfnamefont {A.}~\bibnamefont {Delfino}}, \bibinfo {author} {\bibfnamefont
  {J.~R.}\ \bibnamefont {Stone}}, \ and\ \bibinfo {author} {\bibfnamefont
  {P.~D.}\ \bibnamefont {Stevenson}},\ }\href {\doibase
  10.1103/PhysRevC.85.035201} {\bibfield  {journal} {\bibinfo  {journal} {Phys.
  Rev. C}\ }\textbf {\bibinfo {volume} {85}},\ \bibinfo {pages} {035201}
  (\bibinfo {year} {2012})}\BibitemShut {NoStop}%
\bibitem [{\citenamefont {Baran}\ \emph {et~al.}(2014)\citenamefont {Baran},
  \citenamefont {Palade}, \citenamefont {Colonna}, \citenamefont {{Di Toro}},
  \citenamefont {Croitoru},\ and\ \citenamefont {Nicolin}}]{Bar2014}%
  \BibitemOpen
  \bibfield  {author} {\bibinfo {author} {\bibfnamefont {V.}~\bibnamefont
  {Baran}}, \bibinfo {author} {\bibfnamefont {D.}~\bibnamefont {Palade}},
  \bibinfo {author} {\bibfnamefont {M.}~\bibnamefont {Colonna}}, \bibinfo
  {author} {\bibfnamefont {M.}~\bibnamefont {{Di Toro}}}, \bibinfo {author}
  {\bibfnamefont {A.}~\bibnamefont {Croitoru}}, \ and\ \bibinfo {author}
  {\bibfnamefont {A.}~\bibnamefont {Nicolin}},\ }\href@noop {} {\  (\bibinfo
  {year} {2014})},\ \bibinfo {note} {arXiv: 1411.6965}\BibitemShut {NoStop}%
\bibitem [{\citenamefont {Inakura}\ \emph {et~al.}(2011)\citenamefont
  {Inakura}, \citenamefont {Nakatsukasa},\ and\ \citenamefont
  {Yabana}}]{INY2011}%
  \BibitemOpen
  \bibfield  {author} {\bibinfo {author} {\bibfnamefont {T.}~\bibnamefont
  {Inakura}}, \bibinfo {author} {\bibfnamefont {T.}~\bibnamefont
  {Nakatsukasa}}, \ and\ \bibinfo {author} {\bibfnamefont {K.}~\bibnamefont
  {Yabana}},\ }\href@noop {} {\bibfield  {journal} {\bibinfo  {journal} {Phys.
  Rev. C}\ }\textbf {\bibinfo {volume} {84}},\ \bibinfo {pages} {021302(R)}
  (\bibinfo {year} {2011})}\BibitemShut {NoStop}%
\bibitem [{\citenamefont {Inakura}\ \emph {et~al.}(2013)\citenamefont
  {Inakura}, \citenamefont {Nakatsukasa},\ and\ \citenamefont
  {Yabana}}]{INY2013}%
  \BibitemOpen
  \bibfield  {author} {\bibinfo {author} {\bibfnamefont {T.}~\bibnamefont
  {Inakura}}, \bibinfo {author} {\bibfnamefont {T.}~\bibnamefont
  {Nakatsukasa}}, \ and\ \bibinfo {author} {\bibfnamefont {K.}~\bibnamefont
  {Yabana}},\ }\href@noop {} {\bibfield  {journal} {\bibinfo  {journal} {Phys.
  Rev. C}\ }\textbf {\bibinfo {volume} {88}},\ \bibinfo {pages} {051305(R)}
  (\bibinfo {year} {2013})}\BibitemShut {NoStop}%
\bibitem [{\citenamefont {Piekarewicz}\ \emph {et~al.}(2012)\citenamefont
  {Piekarewicz}, \citenamefont {Agrawal}, \citenamefont {Col{\`o}},
  \citenamefont {Nazarewicz}, \citenamefont {Paar}, \citenamefont {Reinhard},
  \citenamefont {Roca-Maza},\ and\ \citenamefont {Vretenar}}]{Pie2012}%
  \BibitemOpen
  \bibfield  {author} {\bibinfo {author} {\bibfnamefont {J.}~\bibnamefont
  {Piekarewicz}}, \bibinfo {author} {\bibfnamefont {B.}~\bibnamefont
  {Agrawal}}, \bibinfo {author} {\bibfnamefont {G.}~\bibnamefont {Col{\`o}}},
  \bibinfo {author} {\bibfnamefont {W.}~\bibnamefont {Nazarewicz}}, \bibinfo
  {author} {\bibfnamefont {N.}~\bibnamefont {Paar}}, \bibinfo {author}
  {\bibfnamefont {P.}~\bibnamefont {Reinhard}}, \bibinfo {author}
  {\bibfnamefont {X.}~\bibnamefont {Roca-Maza}}, \ and\ \bibinfo {author}
  {\bibfnamefont {D.}~\bibnamefont {Vretenar}},\ }\href@noop {} {\bibfield
  {journal} {\bibinfo  {journal} {Phys. Rev. C}\ }\textbf {\bibinfo {volume}
  {85}},\ \bibinfo {pages} {041302(R)} (\bibinfo {year} {2012})}\BibitemShut
  {NoStop}%
\bibitem [{\citenamefont {Bacca}\ \emph {et~al.}(2014)\citenamefont {Bacca},
  \citenamefont {Barnea}, \citenamefont {Hagen}, \citenamefont {Miorelli},
  \citenamefont {Orlandini},\ and\ \citenamefont {Papenbrock}}]{Bac2014}%
  \BibitemOpen
  \bibfield  {author} {\bibinfo {author} {\bibfnamefont {S.}~\bibnamefont
  {Bacca}}, \bibinfo {author} {\bibfnamefont {N.}~\bibnamefont {Barnea}},
  \bibinfo {author} {\bibfnamefont {G.}~\bibnamefont {Hagen}}, \bibinfo
  {author} {\bibfnamefont {M.}~\bibnamefont {Miorelli}}, \bibinfo {author}
  {\bibfnamefont {G.}~\bibnamefont {Orlandini}}, \ and\ \bibinfo {author}
  {\bibfnamefont {T.}~\bibnamefont {Papenbrock}},\ }\href {\doibase
  10.1103/PhysRevC.90.064619} {\bibfield  {journal} {\bibinfo  {journal} {Phys.
  Rev. C}\ }\textbf {\bibinfo {volume} {90}},\ \bibinfo {pages} {064619}
  (\bibinfo {year} {2014})}\BibitemShut {NoStop}%
\end{thebibliography}
\end{document}